\documentclass[reprint,prl,twocolumn,superscriptaddress,showpacs,showkeys,floatfix,preprintnumbers]{revtex4-1}
\usepackage{braket}
\usepackage{xargs}
\usepackage{mathtools}
\usepackage[english]{babel}
\usepackage[utf8]{inputenc}
\usepackage{diagbox}
\usepackage{amsmath,amssymb,braket,slashed,mathrsfs}
\usepackage{pifont}
\usepackage{MnSymbol}
\usepackage{graphicx}
\graphicspath{{./fig/}}
\usepackage{tikz}
\usetikzlibrary{shapes,arrows,positioning}
\tikzset{decision/.style={diamond, draw, fill=blue!20, text width=4.5em, text badly centered, inner sep=0pt}}
\tikzset{block/.style={rectangle, draw, fill=blue!20, text width=10em, text centered, rounded corners, minimum width=3.5cm}}
\tikzset{block1/.style={rectangle, draw, fill=blue!20, text width=18.5em, text centered, rounded corners, minimum width=3.5cm}}
\tikzset{line/.style={draw, -latex, thick}}
\usepackage{color,hyperref}
\usepackage{multirow,array}
\usepackage{textcomp}
\usepackage{amsmath}

\usepackage{cleveref}

\newcommand{\innovation}{Collaborative Innovation Center of Quantum Matter, Beijing 100871, China}
\newcommand{\chep}{Center for High Energy Physics, Peking University, Beijing 100871, China}
\newcommand{\pkuphy}{School of Physics, Peking University, Beijing 100871,
China}

\newcommand{\Uconn}{Department of Physics, University of Connecticut, Storrs, CT 06269, USA}
\newcommand{\RBRC}{RIKEN-BNL Research Center, Brookhaven National Laboratory, Building 510, Upton, NY 11973}

\newcommand{\itp}{CAS Key Laboratory of Theoretical Physics, Institute of Theoretical Physics,
Chinese Academy of Sciences, Beijing 100190, China}

\newcommand{\nn}{\nonumber}
\newcommand{\ba}{\begin{eqnarray}}
\newcommand{\ea}{\end{eqnarray}}
\newcommand{\be}{\begin{equation}}
\newcommand{\ee}{\end{equation}}

\begin{document}
\title{Nucleon electric polarizabilities and nucleon-pion scattering at the physical pion mass}

\author{Xuan-He~Wang}\affiliation{\pkuphy}
\author{Zhao-Long~Zhang}\affiliation{\pkuphy}
\author{Xiong-Hui~Cao}\affiliation{\itp}
\author{Cong-Ling~Fan}\affiliation{\pkuphy}
\author{Xu~Feng}\email{xu.feng@pku.edu.cn}\affiliation{\pkuphy}\affiliation{\innovation}\affiliation{\chep}
\author{Yu-Sheng~Gao}\affiliation{\pkuphy}
\author{Lu-Chang~Jin}\email{ljin.luchang@gmail.com}\affiliation{\Uconn}\affiliation{\RBRC}
\author{Chuan~Liu}\affiliation{\pkuphy}\affiliation{\innovation}\affiliation{\chep}

%\pacs{PACS}
%

\date{\today}

\begin{abstract}
We present a lattice QCD calculation of the nucleon electric polarizabilities at the physical pion mass.
Our findings reveal the substantial contributions of the $N\pi$ states to these polarizabilities. 
Without considering these contributions, the lattice results fall significantly below the experimental values, consistent with previous lattice studies.
This observation has motivated us to compute both the parity-negative $N\pi$ scattering up to a nucleon momentum of $\sim0.5$ GeV in the center-of-mass frame 
and corresponding $N\gamma^*\to N\pi$ matrix elements using lattice QCD. Our results confirm that incorporating dynamic
$N\pi$ contributions is crucial for a reliable determination of the polarizabilities from lattice QCD. This methodology lays the groundwork for future lattice QCD investigations into various other polarizabilities.
\end{abstract}

\maketitle
\section{Introduction}

As building blocks of the visible universe, 
nucleons play a pivotal role in our pursuit of understanding the internal structure of matter and the unraveling of the mysteries of universe.
Being a bound state of quarks and gluons, nucleons exhibit a complex structure that poses significant challenges to understand, 
particularly at lower energy levels where non-perturbative strong interactions and confinement effects come into play. 
In the realm of nucleon properties, the electric and magnetic polarizabilities represent crucial fundamental constants akin to the size and shape of the proton. 
These polarizabilities, denoted as $\alpha_E$ and $\beta_M$, respectively, offer insights into the distribution of charge and magnetism within nucleons, revealing their response to external electromagnetic fields. 
Experimental determination of polarizabilities relies on processes such as Compton scattering, wherein external electromagnetic fields polarize the target nucleon or deuteron. 

\begin{figure}[htb]
\centering
\includegraphics[width=0.48\textwidth,angle=0]{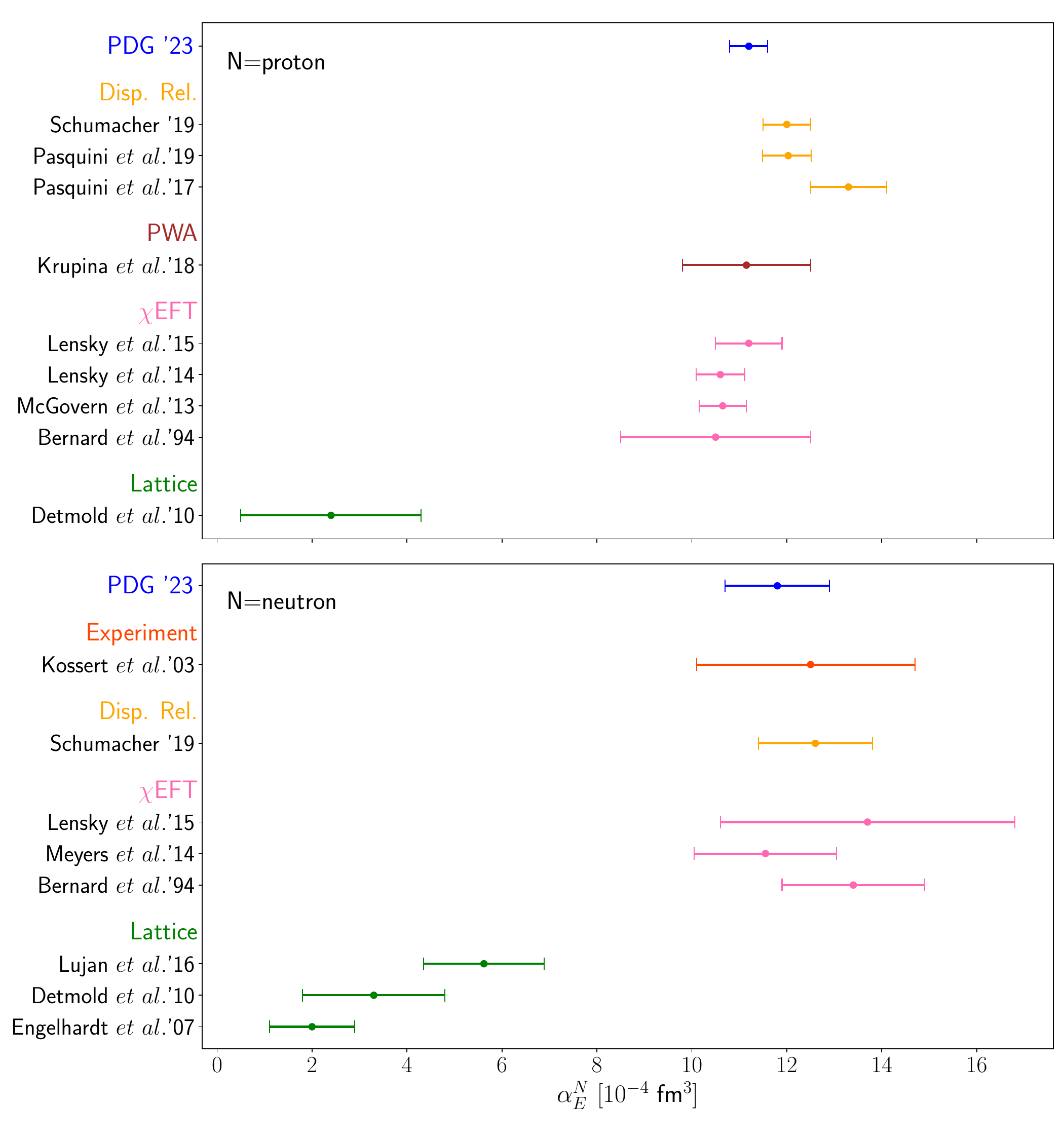}
\caption{
Summary of the electric dipole polarizability for the proton, $\alpha_E^{p}$ (upper panel), and the neutron, $\alpha_E^{n}$ (lower panel). 
We utilize a similar figure as presented in Ref.~\cite{Hagelstein:2020vog}, but with the inclusion of updated PDG data points.
}
\label{fig:summary}
\end{figure}

Traditional perturbative QCD techniques are inadequate at low energies, where strong coupling dominates. Lattice QCD, along with data-driven analysis and effective field theories like chiral perturbation theory ($\chi$PT)~\cite{Bernard:1991rq,Bernard:1991ru,Bernard:1992qa}, offers an avenue to understand nucleon polarizabilities. During the initial stages, lattice QCD calculations of hadron polarizabilities were conducted under the quenched approximation~\cite{Fiebig:1988en,Wilcox:1996vx,Wilcox:1997ee,Christensen:2004ca,Lee:2005dq}. 
However, these calculations couldn't capture the crucial aspect of chiral dynamics, where the nucleon core is surrounded by a pion cloud.
Through the decades of dedicated work, lattice QCD can calculate the hadron polarizabilities with full QCD simulations~\cite{Engelhardt:2007ub,Detmold:2010ts,Lujan:2014kia,Bignell:2018acn,Bignell:2020xkf,Detmold:2009dx,Freeman:2014kka,Bignell:2020dze,He:2021eha,Niyazi:2021jrz,Lee:2023rmz,Lee:2023lnx}. 
A recent lattice study focusing on pion electric polarizabilities~\cite{Feng:2022rkr}
achieved calculations at the physical pion mass ($M_\pi$) for the first time, yielding results comparable to predictions from $\chi$PT.
However, in the case of nucleon $\alpha_E$, lattice results derived from dynamic simulations at heavier-than-physical $M_\pi$~\cite{Engelhardt:2007ub,Detmold:2010ts,Lujan:2014kia} notably deviate from outcomes obtained through data-driven analysis~\cite{Schumacher:2019ikn,Pasquini:2019nnx,Krupina:2017pgr,Pasquini:2017ehj,Kossert:2002ws}, $\chi$EFT~\cite{Lensky:2015awa,Lensky:2014efa,COMPTONMAX-lab:2014cve,McGovern:2012ew,Bernard:1993ry}, or the PDG average~\cite{Workman:2022ynf}, as depicted in Fig.~\ref{fig:summary}. It's worth noting that the lattice QCD computation of nucleon electric polarizabilities represents an initial stride toward tackling various other polarizabilities. Hence, comprehending the origins of systematic effects and advancing toward a benchmark calculation holds paramount importance.

In this work, we extract $\alpha_E$ by computing nucleon four-point correlation (4pt) functions at the physical $M_\pi$.
Upon delving into the detailed contributions from intermediate states in the calculation of these 4pt functions, we discern the pivotal role played by the nucleon-pion ($N\pi$) states. Consequently, we proceed to directly calculate the matrix elements involving $N\pi$, encompassing both isospin $I=\frac{1}{2}$ and $I=\frac{3}{2}$ channels.
These matrix elements allows an accurate, indirect determination of the nucleon 4pt function at long distances.
The results for $\alpha_E^p$ and $\alpha_E^n$ receive substantial contributions from these $N\pi$ states, which would be extremely challenging to obtain through the direct calculation of the 4pt functions alone.
By including the $N\pi$ contributions, our final lattice results agree well with the PDG values.

It is noteworthy that $N\pi$ scattering holds intrinsic significance for the investigation of nucleon interactions and the lowest-lying $N\pi$ resonances. 
While experimental and phenomenological approaches such as the Roy-Steiner equations~\cite{Baru:2010xn,Baru:2011bw,Hoferichter:2015hva,RuizdeElvira:2017stg,Hoferichter:2023ptl} have yielded a comprehension of the scattering scenario, the direct determination of amplitudes from QCD is constrained by its non-perturbative character at low energies, necessitating input from lattice QCD.
Due to the complicated quark field contractions and substantial signal-to-noise challenges, so far,
only a limited number of lattice QCD studies address this subject~\cite{Bulava:2022vpq,Andersen:2017una,Lang:2016hnn,Lang:2012db,Fukugita:1994ve}. 
Until very recently, the first calculation of $I=\frac{3}{2}$ $N\pi$ scattering length ($a_0^I$) at the physical $M_\pi$ was reported by the ETMC collaboration~\cite{Alexandrou:2023elk}. 
In the same vein, we present lattice results for $a_0^I$ ($I=\frac{1}{2},\frac{3}{2}$) at the physical point, utilizing a different QCD discretization scheme.

\section{Lattice methodology}

We begin with the spin-averaged forward doubly-virtual Compton scattering tensor defined in Euclidean space
	\ba
	 \label{eq:Compton_tensor}
	 T^{\mu\nu}(P,Q)&=&\frac{1}{2}\int d^4x\,e^{-iQ\cdot x}\langle N(P)|\mathcal{T}[J^\mu(x)J^\nu(0)]|N(P)\rangle
	\nn\\
	&=&\mathcal{K}_1^{\mu\nu}T_1(Q_0,Q^2)+\mathcal{K}_2^{\mu\nu}T_2(Q_0,Q^2)
	 \ea
where $P= (iM,\vec{0})$ and $Q=(Q_0,\vec{Q})$ represent the Euclidean four-momenta of nucleon and photon, respectively, with $M$ denoting the nucleon mass. $J^{\mu,\nu}$ represent the electromagnetic quark currents, while
$\mathcal{K}_{1,2}^{\mu\nu}$ signify two conserved Lorentz tensors
	\ba
	&&\mathcal{K}_1^{\mu\nu}=\delta^{\mu\nu}Q^2-Q^\mu Q^\nu
	\nn\\
	&&\mathcal{K}_2^{\mu\nu}=(P^\mu Q^\nu+P^\nu Q^\mu)\frac{P\cdot Q}{M^2}-\delta^{\mu\nu}\frac{(P\cdot Q)^2}{M^2}-\frac{P^\mu P^\nu Q^2}{M^2}.
	\nn\\
	\ea
The scalar function $T_i$ can be expressed as a combination of Born ($T_i^B$) and non-Born ($T_i^{NB}$) terms: $T_i=T_i^B+T_i^{NB}$.
The Born terms, derived from the elastic box and crossed box diagrams, encompass the pole generated by the elastic intermediate states~\cite{Birse:2012eb}. In contrast, the non-Born term remains regular as $Q\to0$ and, as a result, can be expanded in a Taylor series involving powers of $Q$. 
The two polarizabilities, $\alpha_E$ and $\beta_M$, determine the leading terms in the low-energy expansion of $T_{i}^{NB}$
\ba
&&T_1^{NB}(Q_0,Q^2)=\frac{M}{\alpha_{em}}[-\beta_M+\mathcal{O}(Q)],
\nn\\
&&T_2^{NB}(Q_0,Q^2)=\frac{M}{\alpha_{em}}[\alpha_E+\beta_M+\mathcal{O}(Q)].
\ea

By conducting a low-$Q$ expansion, we can extract $\alpha_E$ from hadronic functions $H^{\mu\nu}(x)$, defined as
\be
H^{\mu\nu}(x)=\langle N(P)|\mathcal{T}[J^\mu(x)J^\nu(0)]|N(P)\rangle.
\ee
Specifically, we have three formulae available to calculate $\alpha_E$~\cite{SM}:
\be
\label{eq:alpha00}
\alpha_E=-\frac{1}{12}\frac{\alpha_{em}}{M}\int d^4x\,\vec{x}^2\left(H^{00}(x)-H^{00}_{GS}(x)\right)+\alpha_E^{r},
\ee
\be
\label{eq:alpha01}
\alpha_E=\frac{1}{4}\frac{\alpha_{em}}{M}\int d^4x\,(t+x_i)^2\left(H^{0i}(x)-H^{0i}_{GS}(x)\right)
+\alpha_E^{r},
\ee
and 
\be
\label{eq:alpha02}
\alpha_E=-\frac{1}{12}\frac{\alpha_{em}}{M}\int d^4x\,t^2\sum_{i}H^{ii}(x)+\alpha_E^{r},
\ee
with $H^{\mu\nu}_{GS}(x)$ representing the ground-state contribution to  $H^{\mu\nu}(x)$, defined as
\be
H_{GS}^{\mu\nu}(x)=\langle N(P)|\mathcal{T}[J^\mu(x)\hat{P}_NJ^\nu(0)]|N(P)\rangle.
\ee
Here $\hat{P}_N$ is a projection operator for the nucleon state
\be
\hat{P}_N=\int \frac{d^3\vec{Q}}{(2\pi)^3} \frac{1}{2E}|N(\vec{Q})\rangle\langle N(\vec{Q})|,\quad E=\sqrt{M^2+\vec{Q}^2}.
\ee
$\alpha_E^r$ accounts for residual effects that arise when the pole structure between the Born term and ground-state contribution cancels out.
It can be expressed as
\be
\alpha_E^{r}=\frac{\alpha_{em}}{M}\left(\frac{G_E^2(0)+\kappa^2}{4M^2}+\frac{G_E(0)\langle r_E^2\rangle}{3}\right),
\ee
where $G_E(0)$ represents the electric form factor at zero momentum transfer, with $G_E^p(0)=1$ and $G_E^n(0)=0$, respectively. $\kappa$ denotes the nucleon anomalous magnetic moment, 
and $\langle r_E^2\rangle$ indicates the squared charge radius.

These position-space formulae~(\ref{eq:alpha00})-(\ref{eq:alpha02}) for $\alpha_E$ are similar to those invented in Ref.~\cite{Feng:2022rkr} for pions.
Ref.~\cite{Wilcox:2021rtt} has derived a momentum-space formula for $\alpha_E$ based on $H^{00}(p)$ at small momentum $p$.

In our calculation, among the three formulae derived, we find $\sum_i H^{ii}(x)$ and Eq.~\eqref{eq:alpha02} provide the best estimate for $\alpha_E$ for two reasons.
First, $\sum_i H^{ii}(x)$ are not enhanced by $\vec x^2$ near the boundary of the lattice, which leads to smaller finite volume error.
Second, it does not receive contribution from the ground state and does not rely on the cancellation between $H^{\mu\nu}(x)$ and $H^{\mu\nu}_{GS}(x)$, which leads to smaller statistical and systematic errors.
Consequently, we have opted to rely solely on 
$\sum_iH^{ii}(x)$ and Eq.~\eqref{eq:alpha02} for the determination of $\alpha_E$. 

We represent the contribution from $\sum_iH^{ii}(x)$ as $\alpha_E^{ii}$, and consequently, $\alpha_E$ can be expressed as $\alpha_E=\alpha_E^{ii}+\alpha_E^r$.
The term $\alpha_E^{ii}$ encompasses contributions from various intermediate states
\ba
\label{eq:contribution_diff_state}
\alpha_E^{ii}&=&-\frac{1}{12}\frac{\alpha_{em}}{M}\int_{-\infty}^\infty dt\,t^2\sum_n A_{n}e^{-(E_n-M)|t|}
\nn\\
&=&-\frac{1}{3}\frac{\alpha_{em}}{M}\sum_n \frac{A_{n}}{(E_n-M)^3}.
\ea
Typically, $A_n$ can be written as $A_n=\sum_i\langle N(\vec{0})|J^i|n\rangle\langle n|J^i|N(\vec{0})\rangle$, where $|n\rangle$ denotes zero-momentum baryonic states with $E_n$ representing their energies.
These states are normalized as $\langle n' | n \rangle =L^3 \delta_{n'n}$.
(When computing the disconnected diagrams, we need to subtract the product of the nucleon correlator and hadronic vacuum polarization function. $A_n$ then receives another type of contribution:
$A_n=-\sum_i\langle N(\vec{0})|N(\vec{0})\rangle \langle 0 | J^i|n\rangle\langle n|J^i| 0 \rangle$, where $|n\rangle$ denotes the states with zero baryon number, with $E_n$ representing their energies plus $M$.)

For the low-lying $N\pi$ states, the contributions to $\alpha_E^{ii}$
are enhanced by a factor of $\frac{1}{\Delta E_n^3}$ with $\Delta E_n=E_n-M$. 
It's worth noting that the lattice data of $H^{ii}(t)=\int d^3\vec{x}\sum_iH^{ii}(t,\vec{x})$ are
notably noisy if the time separation $t$ between the two currents exceeds 1 fm, due to the signal-to-noise problem. Therefore, it becomes imperative to introduce a temporal truncation denoted as $t_0$ for the integral.
Conversely, the integrand $t^2 e^{-\Delta E t}$ exhibits a peak at $t=\frac{2}{\Delta E_n}$. Particularly, for the lowest
state $|n=0\rangle$, which corresponds to $N\pi$ scattering state
near threshold, the peak is located at $t=\frac{2}{M_\pi}\approx 2.8$ fm, indicating significant temporal truncation effects. 
Given the inputs of $E_n$, $A_n$ and $t_0$, the truncation effects are anticipated as $-\frac{1}{3}\frac{\alpha_{em}}{M}\frac{A_n}{\Delta E_n^3}e^{-\Delta E_n t_0}f(\Delta E_n,t_0)$, where $f(\Delta E_n,t_0)=\frac{1}{2}(\Delta E_nt_0)^2+\Delta E_nt_0+1$.

To address this issue, we compute the $N\pi$ matrix elements for various nucleon and pion momentum modes $\vec{p}=\frac{2\pi}{L}\vec{m}$ with $\vec{m}$ taking integer values and satisfying $|\vec{m}|\le \sqrt{3}$~\cite{SM}. 
For $|\vec{m}|\ge 2$, the temporal truncation effects are suppressed by a factor of $\frac{\Delta E_0^3}{\Delta E_n^3}e^{-\Delta E_n t_0}f(\Delta E_n,t_0)$, which is $\sim4\times10^{-3}$ when taking $t_0=0.75$ fm and can be safely neglected.
With the inputs of low-lying $N\pi$ states, $\alpha_E^{ii}$ can be partitioned into low-lying $N\pi$-state contribution $\alpha_E^{ii,N\pi}$ and the contribution from other higher excited states, denoted as $\alpha_E^{ii,es}$. As $E_n+E_\pi-M$ are always positive, based on the Poisson summation formula,
the finite-volume effects associated with the momentum summation in Eq.~(\ref{eq:contribution_diff_state}) are exponentially suppressed as the lattice size $L$ increases. However, we find that 
at a lattice size of $L\approx 4.6$ fm, significant finite-volume effects still persist. Therefore, we use a momentum integral instead of a momentum summation to compute $\alpha_E^{ii,N\pi}$. The residual
finite-volume effects are estimated to be at the sub-percent level~\cite{SM}.

\section{Numerical analysis}

We utilized two lattice QCD gauge ensembles at the physical $M_\pi$, denoted as 24D and 32Dfine, which feature nearly identical spatial size
 ($L=4.63$ fm and 4.58 fm) but have different lattice spacings ($a^{-1}=1.023(2)$ GeV and 1.378(5) GeV). These ensembles were generated
by RBC and UKQCD Collaborations utilizing $2+1$-flavor domain wall fermion~\cite{RBC:2014ntl}. Further ensemble
parameters can be found in Ref.~\cite{Ma:2023kfr}. We use 207 configurations for 24D and 82 configurations for 32Dfine. 
For each configuration, we make use of 1024 point-source and 1024 smeared-point-source propagators at random spatial-temporal locations
and subsequently computed three types of correlation functions
\ba
&&C_{NJJN}(t_f,t_x,t_y,t_i)=\langle O_{N}(t_f)J_\mu(t_x)J_\nu(t_y)O_{N}^\dagger(t_i)\rangle,
\nn\\
&&C_{NJN\pi}^{I,I'}(t_f,t,t_i)=\langle O_{N\pi}^{I}(t_f)J_\mu^{I'}(t)O_{N}^\dagger(t_i)\rangle,
\nn\\
&&C_{N\pi}^{I}(t_f,t_i)=\langle O_{N\pi}^{I}(t_f){O_{N\pi}^{I}}^\dagger(t_i)\rangle,
\ea
using the random sparsening-field technique~\cite{Li:2020hbj,Detmold:2019fbk}. Here we have performed zero-momentum projection to operators $O_N$, $O_{N\pi}$ and $J^\mu$. Regarding the composite operator $O_{N\pi}$, the $N$ and $\pi$ fields are projected onto opposite momenta.
We decompose the electromagnetic current into isovector and isoscalar
components: $J_\mu=J_\mu^{I=1}+J_\mu^{I=0}$.
The superscripts $I$ and $I'$ denote the isospin.
We utilized two parameters $\Delta t_i$ and $\Delta t_f$ to denote the time separation between the nucleon operator and the current insertion. 
For the correlation function involving two current insertions, $t_{i/f}$ are chosen as
$t_f=\operatorname{max}\{t_x,t_y\}+\Delta t_f$ and
$t_i=\operatorname{min}\{t_x,t_y\}-\Delta t_i$. Local vector current operators 
were contracted with the renormalization factors quoted from Ref.~\cite{Feng:2021zek}.
We calculated 
both connected and disconnected diagrams, discarding only disconnected ones which vanish under the flavor SU(3) limit~\cite{SM}. 
For future studies, it would be desirable to include all the diagrams.

\begin{figure}[htb]
\centering
\includegraphics[width=0.48\textwidth,angle=0]{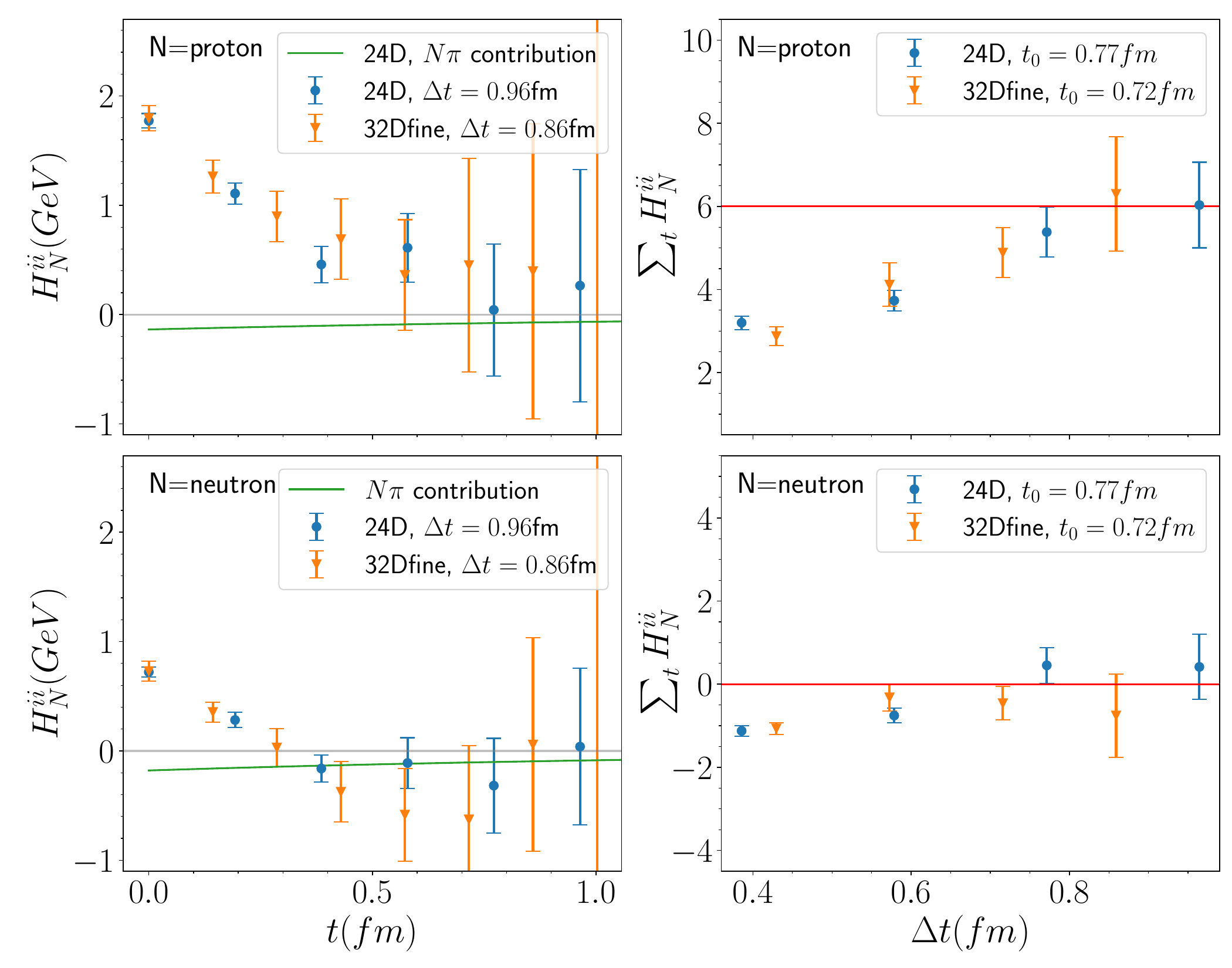}
\caption{$H^{ii}(t)$ for both proton and neutron. 
In the left panel,  we showcase $H^{ii}(t)$ as a function of $t$, while maintaining $\Delta t\simeq0.91$ fm. 
These results are also juxtaposed with contributions from the $N\pi$ ground state. 
The right panel presents the temporal integral $\int_{-t_0}^{t_0} dt\, H^{ii}(t)$ as a function of $\Delta t$.
We employ lattice data at $\Delta t_i=\Delta t_f$ or the averaged result at $\Delta t_i=\Delta t_f\pm a$, 
depending on the value of $\Delta t$.
 At $\Delta t\simeq$ 0.91 fm, the lattice results 
align with the expected values of $K=6\,G_E^2(0)$.
}
\label{fig:Hii}
\end{figure}

In the left panel of Fig.~\ref{fig:Hii}, we present $H^{ii}(t)$ with $\Delta t= \Delta t_i+\Delta t_f$ fixed at $\sim0.96$ fm and $0.86$ fm for 24D and 32Dfine, respectively. 
Clear signals are observed at small $t$. However, as $t$ increases, the signal exhibits an exponential decrease, and noise levels rise rapidly.
Consequently, the lattice data already converge towards 0 at about $0.75$ fm. Therefore, we set the temporal truncation $t_0$ to $0.75$ fm.
According to the current conservation, the low-momentum expansion of the Compton tensor yields $K\equiv \int_{-\infty}^\infty dt\,H^{ii}(t)=6\,G_E^2(0)$~\cite{Fu:2022fgh}.
In the right panel, we present the temporal integral 
$K(t_0)\equiv\int_{-t_0}^{t_0} dt \,H^{ii}(t)$ with $t_0\simeq 0.75$ fm as a function of $\Delta t$. As $\Delta t$ increases, the statistical uncertainties escalate significantly. At 
$\Delta t\simeq 0.91$ fm, we confirm the expected values of $K$, albeit with relatively large uncertainties.
These agreements suggest that the temporal truncation effects are not statistically discernible in the computation of $K$.
It's worth noting that, with $\Delta t\simeq 0.91$ fm and $t_0\simeq 0.75$ fm, the total source-sink separation amounts to 1.6-1.8 fm. This places substantial demands 
on the lattice QCD computation of nucleon 4pt functions. The results presented in Fig.~\ref{fig:Hii} are obtained from connected diagrams. As for those arising from disconnected diagrams,
we observe that $H^{ii,disc}(t)$ generally tend towards zero but yield uncertainties in $\alpha_E^{ii,disc}$ that are 2-3 times larger than the uncertainties of the connected contributions.

\begin{figure}[htb]
\centering
\includegraphics[width=0.48\textwidth,angle=0]{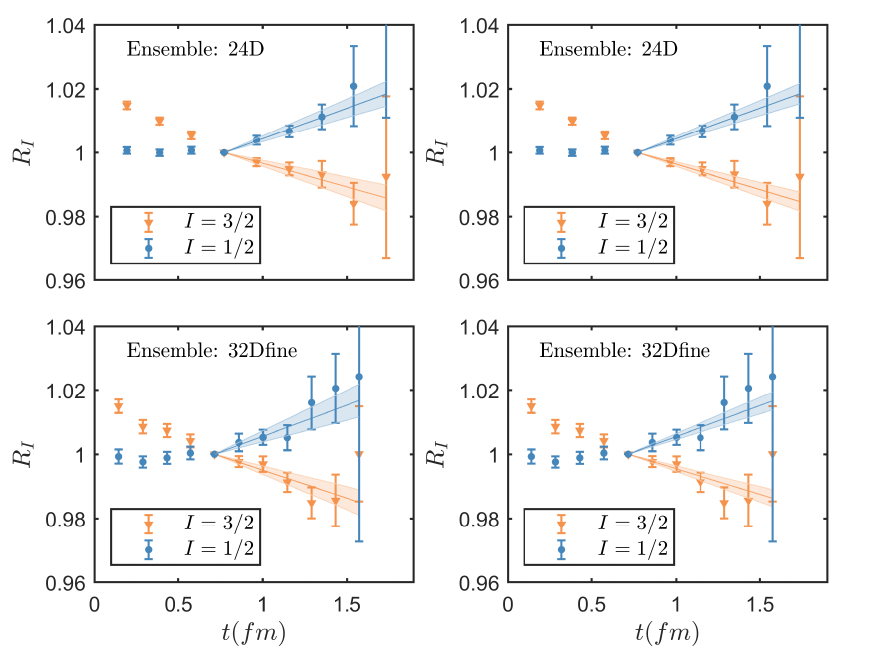}
\caption{$R_I(t)$ for the ground state (after GEVP) as a function of $t$. The 24D and 32Dfine results are depicted in the upper and lower panels, respectively. In the left panel,
we perform the correlated fits to the linear form individually, while in the right panel, we conduct a combined fit using data from both the 24D and 32Dfine ensembles.
}
\label{fig:ratio}
\end{figure}

In the left panel of Fig.~\ref{fig:Hii}, the hadronic functions $H^{ii}(t)$ are contrasted with the contributions from the $N\pi$ ground state. 
These contributions are computed using 
$C_{NJN\pi}^{I,I'}(t_f,t,t_i)$ and $C_{N\pi}^{I}(t_f,t_i)$.
To begin, we calculate the energy shift $\delta E_I$ for the
 $N\pi$ scattering states in the isospin $I=\frac{1}{2}$ and $\frac{3}{2}$ channels by defining the ratio
\be
\label{eq:ratio_energy_shift}
R^I(t)=\frac{C_{N\pi}^I(t_f,t_i)}{C_N(t_f,t_i)C_\pi(t_f,t_i)}
\approx A_I(1-\delta E_I\,t),
\ee
where $t=t_f-t_i$ and $\delta E_I=E_I-M-M_\pi$ with $E_I$ representing the energy for the $N\pi$ state near the threshold.
$C_N$ and $C_\pi$ represent the nucleon and pion two-point correlation functions.
To better isolate the ground state, we also apply the generalized eigenvalue problem (GEVP) method~\cite{Luscher:1990ck,Blossier:2009kd} using the $N\pi$ interpolating operators with 4 lowest momenta. The reference time for GEVP is set at $\sim0.75$ fm.
We denote the method of using Eq.~(\ref{eq:ratio_energy_shift}) as ``before GEVP'' and the other one as ``after GEVP''.
Although before GEVP, the lattice results have already exhibited a linear dependence on $t$ with a reasonable $\chi^2/\mathrm{dof}$ in the fit, after GEVP, we still observe a 0.4-1.2 $\sigma$ downward shift of $\delta E_I$ for various isospin channels and ensembles. As $\delta E_I$ is a tiny quantity, it is not surprising that even for the extraction of the ground states, the results are still sensitive to their mixing with excited states.
In Fig.~\ref{fig:ratio}, we present the $t$ dependence of $R_I(t)$ for the ground state (after GEVP) along with a correlated fit to a linear function of $t$.
Via this fit, we extract $\delta E_I$ for both isospin channels. Remarkably, as the lattice sizes of 24D and 32Dfine are nearly identical (with only a 1\% difference), 
the two ensembles yield highly consistent results for $\delta E_I$, despite their differing lattice spacings. In a prior study of $\pi\pi$ scattering using the same two ensembles~\cite{RBC:2023xqv}, 
$\delta E_I$ for $I=0$ and $2$ exhibited only a 1\% difference between different ensembles. This is not surprising, as 
lattice artifacts are proportionally related to the weak interaction in $\pi\pi$ and $N\pi$ scattering. Furthermore, $\chi$PT informs us that interactions involving $\pi$ at the threshold vanish in the chiral limit.
Therefore, in the $N\pi$ system, we believe that the difference in $\delta E_I$ between 24D and 32Dfine is mainly attributable to the statistical fluctuations rather than lattice artifacts. Instead of conducting the continuum extrapolation, 
we opt for a combined fit using both ensembles.

\begin{figure}[htb]
\centering
\includegraphics[width=0.48\textwidth,angle=0]{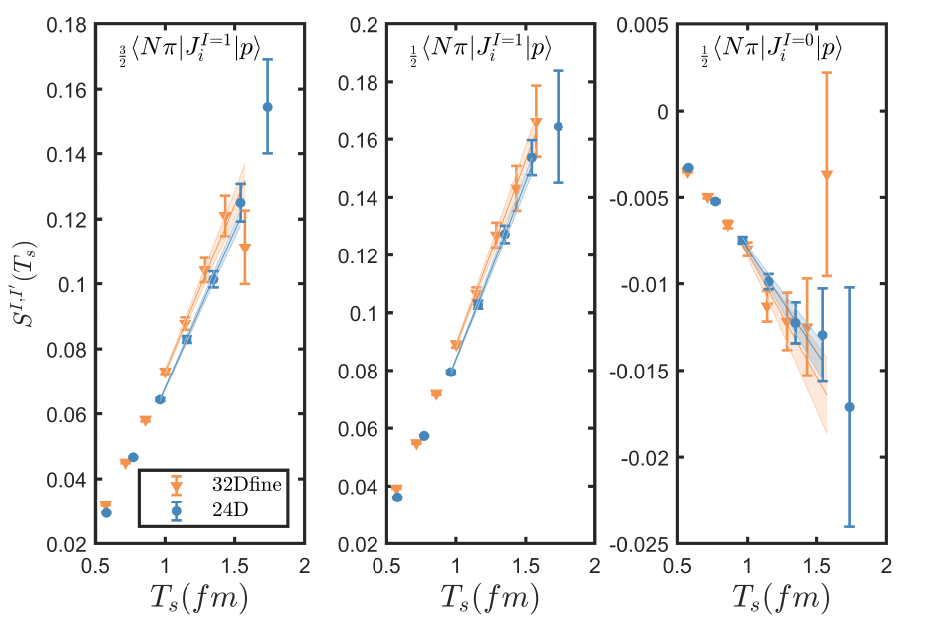}
\caption{The summed insertion $S^{I,I'}(T_s)$ as a function of $T_s$. The three panels showcase $S^{I,I'}(T_s)$ for $\{I,I'\}=\{\frac{3}{2},1\}$, $\{\frac{1}{2},1\}$ and $\{\frac{1}{2},0\}$, respectively.
}
\label{fig:summation_method}
\end{figure}

Next, we compute three types of matrix elements ${_I}\langle N\pi|J_i|N\rangle$: ${_{\frac{3}{2}}}\langle N\pi|J_i^{I=1}|N\rangle$, $_{\frac{1}{2}}\langle N\pi|J_i^{I=1}|N\rangle$ and ${_{\frac{1}{2}}}\langle N\pi|J_i^{I=0}|N\rangle$, with both initial and final states at rest. These matrix elements are normalized as $\langle N|N\rangle = 2ML^3$ and ${_I}\langle N\pi|N\pi\rangle_{I}=L^3$.
We use the ratio 
\ba
R^{I,I'}(t,T_s)&=&\frac{C_{NJN\pi}^{I,I'}(t_f,t,t_i)}{C_{N\pi}^I(t_f,t_i)}
\nn\\
&\times&\sqrt{\frac{C_{N}(t_f,t)C_{N\pi}^I(t,t_i)C_{N\pi}^I(t_f,t_i)}{C_{N\pi}^I(t_f,t)C_N(t,t_i)C_N(t_f,t_i)}}
\ea
with $T_s=t_f-t_i$
to build the summed insertion~\cite{Maiani:1987by,Gusken:1989ad,Bulava:2011yz,Capitani:2012gj}
\ba
S^{I,I'}(T_s)&=&\sum_{t=t_i+a}^{t_f-a}R^{I,I'}(t,T_s)
\nn\\
&\xrightarrow[]{T_s\to\infty} &c_0+\frac{1}{\sqrt{2M}}{_I}\langle N\pi|J_i^{I'}|N\rangle\cdot T_s+\cdots,
\ea
where the dots stand for excited-state contamination that are exponentially suppressed as $T_s$ increases.
By fitting $S^{I,I'}(T_s)$ to a linear form of $T_s$, we extract the dimensionless quantity $\frac{1}{\sqrt{2M}}{_I}\langle N\pi|J_i^{I'}|N\rangle$, as depicted in Fig.~\ref{fig:summation_method}. 
The transition amplitude computed in a finite volume can be converted into the physical amplitude in an infinite volume using Lellouch-L\"uscher factor~\cite{Lellouch:2000pv} and its large-$L$ expansion at the threshold~\cite{Feng:2018pdq}.
The rescattering effects only amount to $\sim10$\%.

For a lattice QCD calculation in a cubic box, the irreducible representation of the cubic group are relevant.
At the $N\pi$ threshold, only the $G_1^-$ representation is involved, while beyond the threshold, both $G_1^-$ and $H^-$ representations come into play. 
The corresponding $N\pi$ operators relevant for $G_1^-$ and $H^-$ representation can be constructed using Eq.~(4.11) in Ref.~\cite{Prelovsek:2016iyo}.

\section{Results and conclusion}

Through the linear fit shown in Fig.~\ref{fig:ratio}, we determine $\delta E_I$ for $I=\frac{1}{2}$ and $\frac{3}{2}$ and extract the scattering length $a_0^I$ (multiplied with $M_\pi$ and denoted as $M_\pi a_0^I$) by incorporating $\delta E_I$ into L\"uscher formula~\cite{Luscher:1986pf,Torok:2009dg}. As depicted in Table~\ref{tab:ensemble_parameter}, the results obtained from individual ensembles are in agreement with each other and also consistent with those from the combined fit. When extracting $M_\pi a_0^I$, we note slight variations in the parameters for $M$, $M_\pi$ and $L$, resulting in a negligible deviation of $0.002$ compared to the statistical error. 
We find that before GEVP, our lattice results exhibit deviations of 1-3$\sigma$ when compared to
the data-driven analysis: $M_\pi a_0^{\frac{1}{2}}=0.170(2)$ and $M_\pi a_0^{\frac{3}{2}}=-0.087(2)$~\cite{Hoferichter:2023ptl}.
However, after GEVP, our results align well with those in Ref.~\cite{Hoferichter:2023ptl}.
Using $M_\pi a_0^I$ as inputs to extract the nucleon sigma term $\sigma_{\pi N}$~\cite{Hoferichter:2015dsa,Hoferichter:2016ocj,Hoferichter:2023ptl}, we obtain $\sigma_{\pi N}=46(28)$ MeV, $34(32)$ MeV and $41(21)$ MeV
for 24D, 32Dfine and the combined fit, respectively.
As $M_\pi a_0^I$ are extracted from $\delta E_I$ at a few MeV level, it is not surprising that the lattice results are much noisier than the results from data-driven analysis.
Within statistical uncertainties, our results for $M_\pi a_0^{\frac{3}{2}}$ also agree with the recent lattice results at the physical $M_\pi$, specifically $M_\pi a_0^{\frac{3}{2}}=-0.13(4)$, as reported by the ETMC collaboration~\cite{Alexandrou:2023elk}.

\begin{table}[htbp]
\small
\centering
\begin{tabular}{c|ccccccc}
\hline
\hline
& Ensemble & 24D & 32Dfine & Combined fit \\
 \hline
& $\chi^2/\mathrm{dof}$ & 0.26 & 0.47 & 0.34 \\
$I=\frac{1}{2}$& $\delta E_I$ [MeV] & $-3.78(79)$ & $-3.9(1.2)$ & $-3.80(66)$ \\
& $M_\pi a_0^I$ & 0.157(37) & 0.157(56) & 0.157(31)\\
\hline
& $\chi^2/\mathrm{dof}$ & 0.26 & 0.51 & 0.38 \\
$I=\frac{3}{2}$ & $\delta E_I$ [MeV] & $2.91(80)$ & $3.40(90)$ & $3.13(60)$ \\
& $M_\pi a_0^I$ & $-0.098(25)$ & $-0.111(27)$ & $-0.104(18)$\\
\hline
\end{tabular}%
\caption{
Fitting results (after GEVP) for the energy shifts ($\delta E_I$) and scattering lengths ($M_\pi a_0^I$) in both $I=\frac{1}{2}$ and $\frac{3}{2}$ channels, obtained from individual ensembles as well as the combined analysis. }
\label{tab:ensemble_parameter}%
\end{table}%

The final results for $\alpha_E^{ii,N\pi}$, $\alpha_E^{ii,es}$, $\alpha_E^{ii,disc}$, $\alpha_E^{r}$, and the total contribution $\alpha_E$ are summarized in Table~\ref{tab:results}. These results show good agreement with the PDG values~\cite{Workman:2022ynf}. Notably, the $N\pi$ states with momentum up to ~0.5 GeV account for about $60\%$ of $\alpha_E^p$ and $90\%$ of $\alpha_E^n$.
This not only helps to partially explain why previous lattice results were significantly lower but also highlights the substantial challenges in lattice QCD calculations of polarizabilities. By computing the 4 lowest $N\pi$ states and replacing momentum summation with momentum integration, both temporal truncation effects and finite-volume effects are reduced to a negligible level. However, since the calculation relies on momentum interpolation, it is advisable to mitigate the uncertainties introduced by interpolation by using larger lattice volumes in future studies. To achieve a precision calculation, contributions from the resonances $N^*/\Delta^*$ and $N\pi$ rescattering effects should also be included.

\begin{table}[htbp]
\small
\centering
\begin{tabular}{c|lcccccc}
\hline
\hline
&  & 24D & 32Dfine & PDG \\
 \hline
 & $\alpha_E^{ii,N\pi}$ & 5.65(53) & 6.5(1.2) & - \\
 & $\alpha_E^{ii,es}$ & $-1.42(53)$ & $-1.58(61)$ & - \\
proton & $\alpha_E^{ii,disc}$ & $1.4(1.2)$ & $0.1(1.8)$ & -\\
 & $\alpha_E^{r}$ & 4.333(3) & 4.333(3) & -\\
 \cline{2-5}
 & $\alpha_E$ & $10.0(1.3)$ & $9.3(2.2)$ & 11.2(4)\\
\hline
& $\alpha_E^{ii,N\pi}$ & 8.33(75) & 9.8(1.5) & - \\
 & $\alpha_E^{ii,es}$ & $-0.60(40)$ & $-0.40(47)$ & - \\
neutron  & $\alpha_E^{ii,disc}$ & $1.4(1.2)$ & $0.1(1.8)$ & -\\
& $\alpha_E^{r}$ & 0.618(1) & 0.618(1) & -\\
\cline{2-5}
&$\alpha_E$ & $9.7(1.4)$ & $10.1(2.4)$ & 11.8(1.1)\\
\hline
\end{tabular}%
\caption{Proton and neutron polarizabilities determined in this work. We list the results of $\alpha_E^{ii,N\pi}$, $\alpha_E^{ii,es}$, $\alpha_E^{ii,disc}$, $\alpha_E^{r}$ and $\alpha_E$ as well as the PDG values in units of $10^{-4}$ fm$^3$.}
\label{tab:results}%
\end{table}%

\begin{acknowledgments}
{\bf Acknowledgments} -- X.F. and L.C.J. gratefully acknowledge many helpful discussions with our colleagues from the
RBC-UKQCD Collaborations. We would like to express our gratitude to Martin Hoferichter for the discussion on extracting the nucleon sigma term from the scattering length, and to De-Liang Yao for the discussion on chiral perturbation theory.
C.L.F., X.F., Y.S.G., C.L., X.H.W. and Z.L.Z. were supported in part by NSFC of China under Grant No. 12125501, No. 12293060, No. 12293063, No. 12070131001, and No. 12141501,
and National Key Research and Development Program of China under No. 2020YFA0406400.
L.C.J. acknowledges support by DOE Office of Science Early Career Award No. DE-SC0021147 and DOE Award No. DE-SC0010339.
X. H.C. was supported in part by NSFC of China under Grant No. 12347120, and the Postdoctoral Fellowship Program of China Postdoctoral Science Foundation under Grant No. GZC20232773 and No. 2023M74360.
The research reported in this work was carried out using the computing facilities at Chinese National Supercomputer Center in Tianjin.
It also made use of computing and long-term storage facilities of the USQCD Collaboration, which are funded by the Office of Science of the U.S. Department of Energy.
\end{acknowledgments}

\bibliography{ref}

\clearpage

\setcounter{page}{1}
\renewcommand{\thepage}{Supplemental Material -- S\arabic{page}}
\setcounter{table}{0}
\renewcommand{\thetable}{S\,\Roman{table}}
\setcounter{equation}{0}
\renewcommand{\theequation}{S\,\arabic{equation}}
\setcounter{figure}{0}
\renewcommand{\thefigure}{S\,\arabic{figure}}

\section{Supplemental Material}

\subsection{Useful formulae to extract $\alpha_E$ and $\beta_M$}

It is not unique to determine $\alpha_E$ and $\beta_M$ using the hadronic functions $H^{\mu\nu}(x)$ as inputs. 
With $P=(iM,\vec{0})$, $Q=(Q_0,\vec{Q})$ as inputs, we have
\ba
&&\mathcal{K}_1^{00}=\mathcal{K}_2^{00}=\vec{Q}^2,\quad \mathcal{K}_1^{0i}=\mathcal{K}_2^{0i}=-Q_0Q_i,
\nn\\
&&\mathcal{K}_1^{ii}=Q_0^2+\vec{Q}^2-Q_i^2,\quad \mathcal{K}_2^{ii}=Q_0^2,
\nn\\
&&\mathcal{K}_1^{ij}=-Q_iQ_j,\quad \mathcal{K}_2^{ij}=0.
\ea
The Born terms are given as follows~\cite{Birse:2012eb}
\ba
\label{eq:T12B}
T_1^B(Q_0,Q^2)&=&-\frac{Q^2(2F_1F_2+F_2^2)+Q_0^2F_2^2}{Q^4+4M^2Q_0^2},
\nn\\
T_2^B(Q_0,Q^2)&=&\frac{Q^2F_2^2+4M^2F_1^2}{Q^4+4M^2Q_0^2}.
\ea	
Here the factor $1/(Q^4+4M^2Q_0^2)=1/[(s-M^2)(u-M^2)]$ emerges from the poles present in the $s$- and $u$-channels with $s=-(P+Q)^2$ and $u=-(P-Q)^2$. In Eq.~(\ref{eq:T12B}), $F_{1,2}(Q^2)$ refer to the nucleon's Dirac
and Pauli form factors, respectively.

The low-$Q$ expansion of the Compton tensors is expressed as
\ba
\label{eq:T_munu}
T^{00}&=&\vec{Q}^2\left(T_1^B+T_2^B+\frac{M}{\alpha_{em}}\alpha_E\right)+\mathcal{O}(Q^3),
\nn\\
T^{0i}&=&-Q_0Q_i\left(T_1^B+T_2^B+\frac{M}{\alpha_{em}}\alpha_E\right)+\mathcal{O}(Q^3),
\nn\\
T^{ii}&=&Q_0^2\left(T_1^B+T_2^B+\frac{M}{\alpha_{em}}\alpha_E\right)
\nn\\
&+&(\vec{Q}^2-Q_i^2)\left(T_1^B-\frac{M}{\alpha_{em}}\beta_M\right)+\mathcal{O}(Q^3)
\nn\\
T^{ij}&=&-Q_iQ_j\left(T_1^B-\frac{M}{\alpha_{em}}\beta_M\right)+\mathcal{O}(Q^3)
\ea
Consequently, we have the flexibility to use either $H^{00}$, $H^{0i}$ or $H^{ii}$ to extract $\alpha_E$ and either $H^{ii}$ or $H^{ij}$ to extract $\beta_M$.

In the extraction of $\alpha_E$, as $T_1^B$ and $T_2^B$ are consistently combined, we define their combination as $T^B$:
\be
\label{eq:T_B}
T^B\equiv T_1^B+T_2^B=\frac{4M^2G_E^2(Q^2)}{Q^4+4M^2Q_0^2}-\frac{F_2^2(Q^2)}{4M^2}.
\ee
Here we have introduced the electric and magnetic form factors
\ba
G_E(Q^2)&=&F_1(Q^2)-\frac{Q^2}{4M^2}F_2(Q^2),
\nn\\
 G_M(Q^2)&=&F_1(Q^2)+F_2(Q^2).
\ea
The low-$Q^2$ expansion of $G_{E/M}$ yields
\ba
G_E(Q^2)&=&G_E(0)-\frac{\langle r_E^2\rangle}{6}Q^2+\frac{\langle r_E^4 \rangle}{120}Q^4+\mathcal{O}(Q^6),
\nn\\
G_M(Q^2)&=&G_M(0)+\mathcal{O}(Q^2)=G_E(0)+\kappa+\mathcal{O}(Q^2),
\nn\\
\ea
where $\langle r_E^2\rangle$ and $\langle r_E^4\rangle$ represent squared and quartic charge radius. $\kappa$ stands for the anomalous magnetic moment of the nucleon.

It's important to note that $T^B$ contains the pole structure, which should be compensated by the ground-state contributions to $H^{\mu\nu}(t,\vec{x})$. These contributions, denoted as $H_{GS}^{\mu\nu}$,
are given by
\ba
H_{GS}^{\mu\nu}(t,\vec{x})\Big|_{t>0}&=& \int \frac{d^3\vec{Q}}{(2\pi)^3}\,\frac{e^{-(E-M)t}e^{i\vec{Q}\cdot\vec{x}}}{2E}
\nn\\
&&
\langle N(\vec{0})|J^\mu(0)|N(\vec{Q})\rangle\langle N(\vec{Q})| J^\nu(0)]|N(\vec{0})\rangle,
\nn\\
H_{GS}^{\mu\nu}(t,\vec{x})\Big|_{t<0}&=&H_{GS}^{\nu\mu}(-t,-\vec{x}),
\ea
where the matrix elements' product is described as
\ba
\mathcal{M}^{\mu\nu}&\equiv&\langle N(\vec{0})|J^\mu(0)|N(\vec{Q})\rangle\langle N(\vec{Q})| J^\nu(0)|N(\vec{0})\rangle
\nn\\
&=&
\begin{cases}
2M(E+M)G_E^2(Q_{on}^2), & \mu=\nu=0,\\ 
-2iMQ_iG_E^2(Q_{on}^2), & \mu,\nu=0,i,\\
-Q_{on}^2\left[\frac{Q_i^2}{\vec{Q}^2}G_E^2+\left(1-\frac{Q_i^2}{\vec{Q}^2}\right)G_M^2\right], & \mu=\nu=i,\\
Q_{on}^2\frac{Q_iQ_j}{\vec{Q}^2}[G_M^2(Q_{on}^2)-G_E^2(Q_{on}^2)],&\mu=i,\nu=j.
\end{cases}
\nn\\
\ea
The squared momentum $Q_{on}^2$ 
results from the momentum transfer between two on-shell states and is represented as
\be
Q_{on}^2=(iE-iM)^2+\vec{Q}^2=2M(E-M).
\ee
Putting the matrix elements' product into the Compton tensor,  we arrive at
\be
\label{eq:T_GS}
T_{GS}^{\mu\nu}=\mathcal{M}^{\mu\nu}\frac{M}{E}\times
\begin{cases}
\frac{Q_{on}^2}{Q_{on}^4+4M^2Q_0^2},& \vec{Q}\neq0,\\
\frac{\sin(Q_0t_0)}{2MQ_0},&\vec{Q}=0,
\end{cases}
\ee
for $\mu=\nu=0$, $\mu=\nu=i$ and $\mu=i,\nu=j$, and
\be
\label{eq:T_GS1}
T_{GS}^{\mu\nu}=\mathcal{M}^{\mu\nu}\frac{M}{E}\times
\begin{cases}
\frac{-i2MQ_0}{Q_{on}^4+4M^2Q_0^2},& \vec{Q}\neq0,\\
\frac{-i(1-\cos(Q_0t_0))}{2MQ_0},& \vec{Q}=0,\\
\end{cases}
\ee
for $\mu,\nu=0,i$.
In the case of $\vec{Q}=0$, a temporal truncation $t_0$ has been introduced.

By combining Eqs.~(\ref{eq:T_munu}), (\ref{eq:T_B}), (\ref{eq:T_GS}) and (\ref{eq:T_GS1}) and considering the momentum assignments,  
$Q=(0,\vec{\xi})$ for $T^{00}$, $(\xi,0,0,\xi)$ for  $T^{03}$ and $(\xi,\vec{0})$ for $T^{ii}$, we can derive three formulae for determining $\alpha_E$, specifically
\ba
\alpha_E&=&\lim_{\vec{Q}^2=\xi^2\to0}
\frac{\alpha_{em}}{M}\left(\frac{T^{00}-T_{GS}^{00}}{\vec{Q}^2}-\frac{\vec{Q}^2T^B-T_{GS}^{00}}{\vec{Q}^2}\right)
\nn\\
&=&-\frac{1}{12}\frac{\alpha_{em}}{M}\int d^4x\,\vec{x}^2\left(H^{00}(x)-H^{00}_{GS}(x)\right)+\alpha_E^{r},
\nn\\
\ea
\ba
\alpha_E&=&\lim_{Q_0^2=Q_3^2=\xi^2\to0}
\frac{\alpha_{em}}{M}\left(\frac{T^{03}-T_{GS}^{03}}{-Q_0Q_3}-\frac{-Q_0Q_3T^B-T_{GS}^{03}}{-Q_0Q_3}\right)
\nn\\
&=&\frac{1}{4}\frac{\alpha_{em}}{M}\int d^4x\,(t+z)^2\left(H^{03}(x)-H^{03}_{GS}(x)\right)
+\alpha_E^{r},
\nn\\
\ea
and
\ba
\alpha_E&=&\lim_{Q_0^2=\xi^2\to0}
\frac{\alpha_{em}}{M}\left(\frac{T^{ii}}{Q_0^2}-T^B\right)
\nn\\
&=&-\frac{1}{4}\frac{\alpha_{em}}{M}\int d^4x\,t^2H^{ii}(x)+\alpha_E^{r},
\ea
with
\be
\alpha_E^{r}=\frac{\alpha_{em}}{M}\left(\frac{G_E^2(0)+\kappa^2}{4M^2}+\frac{G_E(0)\langle r_E^2\rangle}{3}\right).
\ee

To calculate $\beta_M$, we can use $Q=(0,\xi,0,0)$ for $T^{33}$ or $Q=(0,0,\xi,\xi)$ for $T^{23}$ and obtain two formulae
\ba
\label{eq:beta_M0}
\beta_M&=&\lim_{\vec{Q}^2=\xi^2\to0}
-\frac{\alpha_{em}}{M}\left(\frac{T^{33}(\vec{\xi})-T_{GS}^{33}(\vec{\xi})-T_{ES}^{33}(\vec{0})}{Q_1^2}\right.
\nn\\
&&\hspace{1cm}\left.-\frac{Q_1^2T_1^B(\vec{\xi})-T_{GS}^{33}(\vec{\xi})-T_{ES}^{33}(\vec{0})}{Q_1^2}\right)
\\
\label{eq:beta_M1}
&=&\frac{1}{4}\frac{\alpha_{em}}{M}\int d^4x\,x^2\left(H^{33}(x)-H_{GS}^{33}(x)\right)
\nn\\
&&-\frac{\alpha_{em}}{M}\left(\frac{G_E^2(0)}{2M^2}+\frac{G_E(0)\langle r_E^2\rangle}{3}\right)+\beta_M^{r},
\ea
and
\ba
\label{eq:beta_M2}
\beta_M&=&\lim_{Q_2=Q_3=\xi\to0}\frac{\alpha_{em}}{M}\left(\frac{T^{23}-T^{23}_{GS}}{Q_2Q_3}-\frac{-Q_2Q_3T_1^B-T^{23}_{GS}}{Q_2Q_3}\right)
\nn\\
&=&-\frac{1}{4}\frac{\alpha_{em}}{M}\int d^4x\,(y+z)^2\left(H^{23}(x)-H^{23}_{GS}(x)\right)+\beta_M^{r},
\nn\\
\ea
where 
\ba
&&T_{ES}^{ii}(\vec{\xi})\equiv T^{ii}(\vec{\xi})-T_{GS}^{ii}(\vec{\xi}),\quad
\nn\\
&&T^{ii}(\vec{0})=G_E^2(0),\quad
T_{GS}^{ii}(\vec{0})=0,
\ea
and
\ba
\beta_M^{r}&=&-\frac{\alpha_{em}}{M}\frac{G_E(0)\kappa+\kappa^2}{2M^2}.
\ea
The momentum-space formula~\eqref{eq:beta_M0} agrees with the corresponding formula in Ref.~\cite{Wilcox:2021rtt}.
In this work, we have also introduced the position-space formulae~\eqref{eq:beta_M1} and~\eqref{eq:beta_M2}, similar to those derived in Ref.~\cite{Feng:2022rkr} for pions.

\subsection{Contributions of the low-lying $N\pi$ states}
In our study, we calculate the finite-volume matrix elements of ${_L}\langle N\pi,I,\Gamma,r,n|J^i|N\rangle$, where the initial states are either the proton or neutron, and the final states are the $N\pi$ states with total isospin $I=\frac{1}{2}$ or $\frac{3}{2}$, irreducible representation $\Gamma=G_1^{-}$ or $H^-$ of the cubic group (with $r=1,\cdots,\operatorname{dim}(\Gamma)$ indicating the basis of $\Gamma$), and $n=0,1,2,3$  representing  4 lowest eigenstates. These states correspond to different momenta $\vec{p}=\frac{2\pi}{L}\vec{m}$ carried by $N(\vec{p})\pi(-\vec{p})$ with $\vec{m}$ taking the integer values. The momentum degeneracy for the states ${_L}\langle N\pi,I,\Gamma,r,n|$ is given by $\nu_n=\{1,6,12,8\}$.
Note that for the $H^-$ representation, the zero-momentum mode does not contribute. By multiplying a conversion factor $\frac{L^3}{\nu_n}$, the squared matrix elements in the finite volume $|_L\langle N\pi,I,\Gamma,r,n|J^i|N\rangle|^2$ can be related to those in the infinite volume
$|_{\infty}\langle N\pi,I,\Gamma,r,n|J^i|N\rangle|^2$. This conversion factor is equivalent to the Lellouch-L\"uscher factor when $N\pi$ rescattering effects are neglected. 

\begin{figure}[htb]
\centering
\includegraphics[width=0.48\textwidth,angle=0]{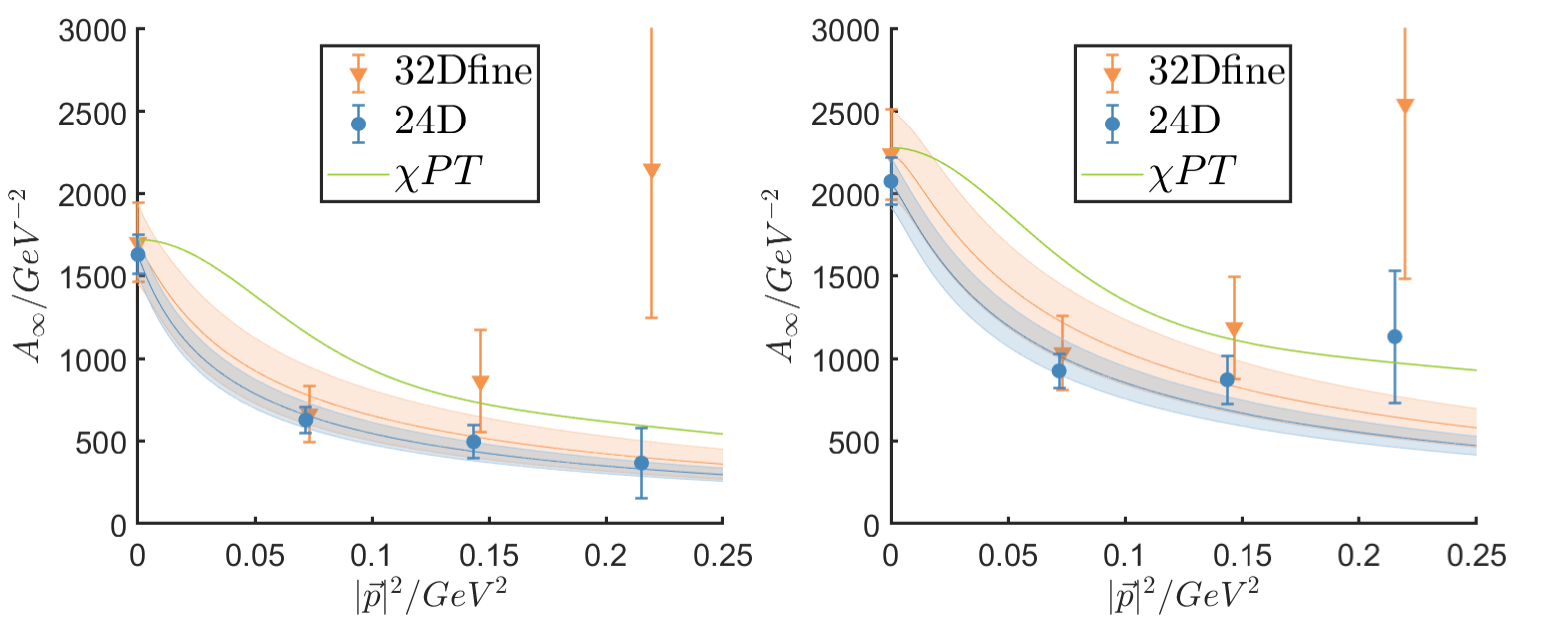}
\caption{$A_\infty$ as a function of squared nucleon momentum $\vec{p}^2$. The left panel shows the results for the proton transition amplitude, and the right panel shows the results for the neutron transition amplitude.
}
\label{fig:fit_pol}
\end{figure}

After obtaining ${_\infty}\langle N\pi,I,\Gamma,r,n|J^i|N\rangle$, we define the amplitude in the infinite volume as
\be
A_\infty=\sum_{I=\frac{1}{2},\frac{3}{2}}\sum_{\Gamma=G_1^-,H^-}\sum_{r=1}^{\operatorname{dim}(\Gamma)}\sum_{i=1,2,3}|_\infty\langle N\pi,I,\Gamma,r,n|J^i|N\rangle|^2
\ee
and use the fit form 
\be
\label{eq:fit_form}
A_\infty=\frac{\sum_s a_s(p^2)^s}{(2E)(2E_\pi)(E+E_\pi-M)^2}
\ee
to describe its momentum dependence. Here, the factors $\frac{1}{2E}$ and $\frac{1}{2E_\pi}$ arise from the conventional normalization condition, and the factor $\frac{1}{(E+E_\pi-M)^2}$ originates from the pole
structure of the transition $N(p_1)+\gamma^*(k)\to\pi(q)+N(p_2)$ in the $s$, $t$, and $u$ channels. Specifically, in our calculation, we have 
$s=(p_2+q)^2=(E_\pi+E)^2$, $u=(p_1-q)^2=(M-E_\pi)^2-\vec{p}^2$ and $t=(p_1-p_2)^2=(M-E)^2-\vec{p}^2$. The associated poles are given by
\ba
&&\frac{1}{s-M^2}=\frac{1}{E_\pi+E+M}\frac{1}{E_\pi+E-M},
\nn\\
&&\frac{1}{u-M^2}=-\frac{1}{M-E_\pi+E}\frac{1}{-M+E_\pi+E},
\nn\\
&&\frac{(q+p_1-p_2)_\mu}{t-M_\pi^2}=-\frac{\delta_{\mu0}}{-M+E+E_\pi}.
\ea
The same pole structure $\frac{1}{(E+E_\pi-M)}$ appears in all $s$, $u$, and $t$ channels.
Note that it is not appropriate to perform a Taylor expansion of $\frac{1}{E+E_\pi-M}$ and $\frac{1}{2E_\pi}$ with respect to momentum $\vec{p}^2$ as the $\vec{p}^2$ and $M_\pi^2$ are of the same order. Therefore, we maintain the denominator as shown in Eq.~(\ref{eq:fit_form}) and apply the Taylor expansion only for the numerator. With limited momentum modes, here we use a linear form for the numerator. The lattice data and the fitting curves are shown in Fig.~\ref{fig:fit_pol}, with the proton transition amplitude in the left panel and the neutron transition amplitude in the right panel. The 24D and 32Dfine results are consistent with each other but are somewhat smaller than the results from next-to-leading-order chiral perturbation theory~\cite{GuerreroNavarro:2020kwb,Cao:2021kvs}.

Once the momentum dependence of $A_\infty$ is determined, we perform an integral to compute
\be
\label{eq:alpha_integral}
\alpha_E^{ii,N\pi}=\frac{1}{3}\frac{\alpha_{em}}{M}\int_{|\vec{p}|<\Lambda}\frac{d^3\vec{p}}{(2\pi)^3}\,\frac{A_\infty}{(E+E_\pi-M)^3},
\ee
with momentum cutoff given by
\be
\label{eq:mom_cutoff}
\Lambda=\left(\frac{2\pi}{L}\right)\left(\left(\sum_{n=0}^3\nu_n\right)/\left(\frac{4}{3}\pi\right)\right)^\frac{1}{3}.
\ee
For $L=4.6$ fm, $\Lambda$ is about 0.5 GeV. Another subtlety is that the matrix element product $\langle N| J^i|N\pi\rangle\langle N\pi|J^i|N\rangle$ differs from the squared matrix element $|\langle N| J^i|N\pi\rangle|^2$ by a minus sign. Therefore, in Eq.~(\ref{eq:alpha_integral}), the coefficient differs from that in Eq.~(\ref{eq:contribution_diff_state}) by a minus sign. 

\subsection{Finite-volume effects}

The finite-volume effects in lattice calculation of polarizabilities are an important issue~\cite{Hall:2013dva,Tiburzi:2014zva}. With our formulation, the finite-volume effects are exponentially suppressed. However, limited by the lattice sizes we used in our calculation, the finite-volume errors remain noticeable. Here, we discuss our method for estimating and correcting these already exponentially suppressed errors.

After determining $A_\infty$, we can estimate the finite-volume effects for the low-lying $N\pi$ states by comparing the discrete momentum summation in the finite volume with the momentum integral in the infinite volume
\ba
\Delta(L)&=&\alpha_E^{ii,N\pi}(L)-\alpha_E^{ii,N\pi}
\nn\\
&=&\frac{1}{3}\frac{\alpha_{em}}{M}\left(\frac{1}{L^3}\sum_{\vec{p}=\frac{2\pi}{L}\vec{m}}-\int\frac{d^3\vec{p}}{(2\pi)^3}\right)\frac{A_\infty}{(E+E_\pi-M)^3},
\nn
\\
\ea
where $\vec{m}$ takes the integer values.

\begin{figure}[htb]
\centering
\includegraphics[width=0.4\textwidth,angle=0]{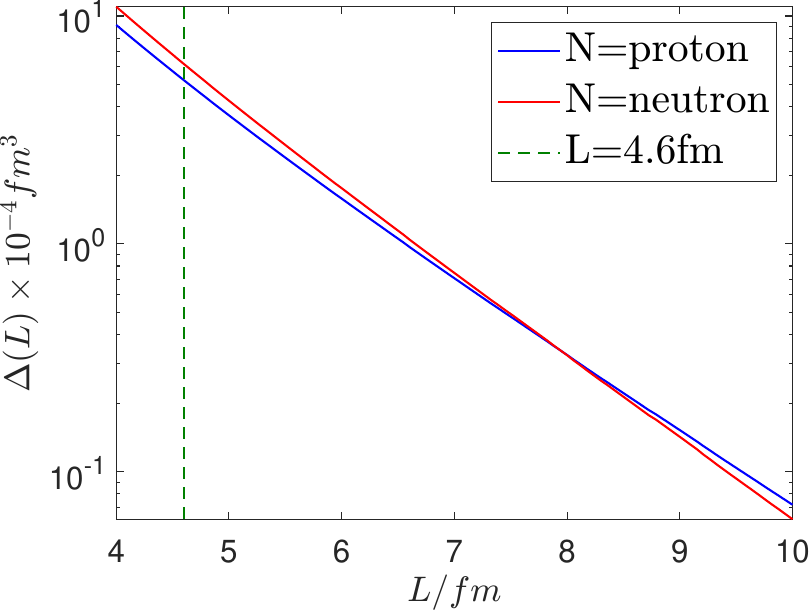}
\caption{The finite-volume effects $\Delta(L)$ as a function of the lattice spatial size $L$.
}
\label{fig:delta_L}
\end{figure}

In Fig.~\ref{fig:delta_L}, we show $\Delta(L)$ as a function of the lattice spatial size $L$. Since there are no singularities in the integrand for any real value of $\vec{p}$, $\Delta(L)$ is exponentially suppressed as $L$ increases, based on Poisson summation formula. This is confirmed by the curves shown in Fig.~\ref{fig:delta_L}. On the other hand, 
the integrand becomes singular in the chiral limit. Consequently, with the lattice size 
used in this calculation, specifically $L\approx4.6\ \text{fm}$, significant finite-volume effects persist.
To reduce the finite-volume effects, we replaced the momentum summation with the momentum integral for low-lying $N\pi$ states up to the momentum cutoff $\Lambda$ defined in Eq.~(\ref{eq:mom_cutoff}). For the residual finite-volume effects from momentum region with $|\vec{p}|>\Lambda$, we estimate them to be $-2.6\times10^{-6}\text{ fm}^3$ for proton and $-4.2\times10^{-6}\text{ fm}^3$ for neutron. Although the form of $A_\infty$ defined in Eq.~(\ref{eq:fit_form}) are not valid for a very large momentum region, these small values still suggest that the residual finite-volume effects are negligible.

\subsection{Contributions from disconnected diagrams}

\begin{figure}[htb]
\centering
\includegraphics[width=0.235\textwidth,angle=0]{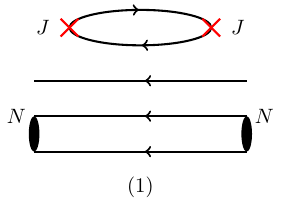}
\includegraphics[width=0.235\textwidth,angle=0]{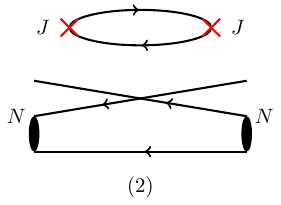}
\includegraphics[width=0.48\textwidth,angle=0]{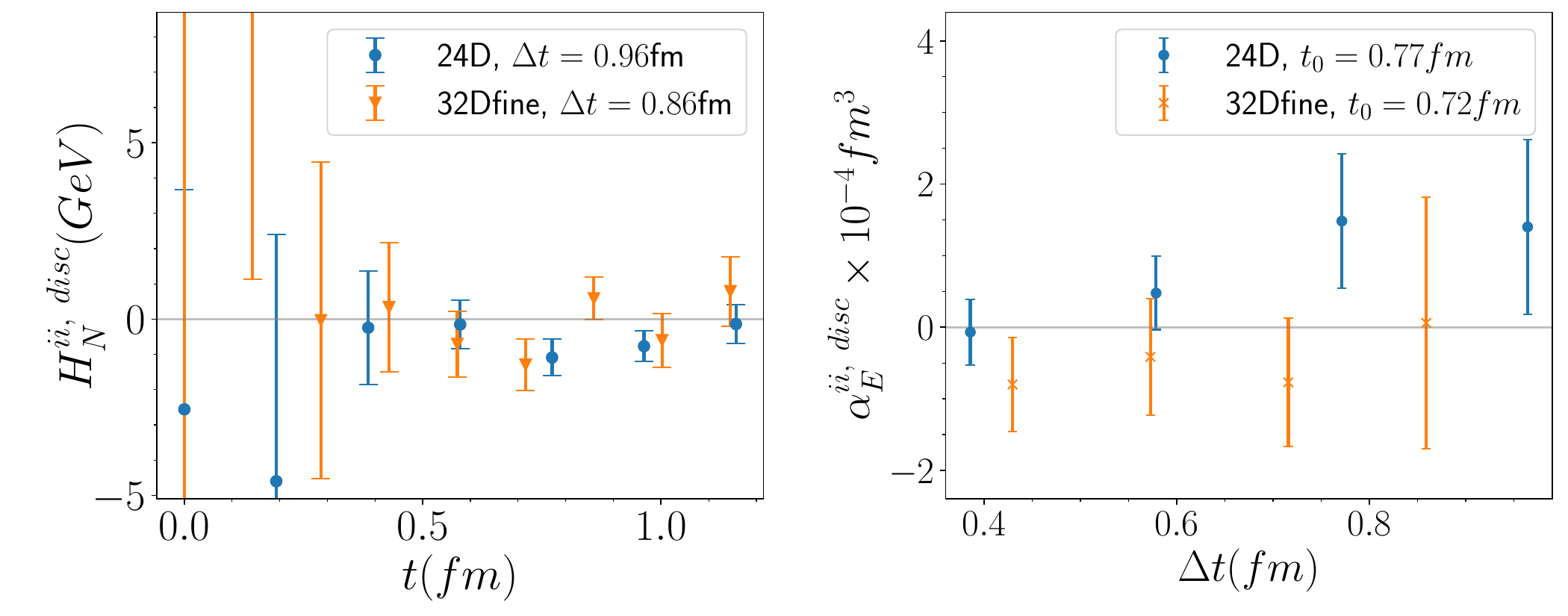}
\caption{Upper panel: two disconnected diagrams computed in this work. Lower panel (left): hadronic functions $H^{ii}(t)$ obtained from disconnected diagrams as a function of $t$ with $\Delta t$ fixed at 0.96 fm (24D) and 0.86 fm (32Dfine). Lower panel (right): the contributions to polarizabilities from disconnected diagrams, $\alpha_E^{ii,disc}$, as a function of $\Delta t$ with the temporal truncation set as $t_0=0.77$ fm (24D) and 0.72 fm (32Dfine). 
}
\label{fig:disc_diag}
\end{figure}

In Fig.~\ref{fig:disc_diag}, we present the results computed from disconnected diagrams. It's worth noting that both proton and neutron $\alpha_E$ share the same disconnected contributions. After performing the vacuum subtraction, we find that the statistical uncertainties for $H^{ii}(t)$ at small $t$ are very large due to the cancellation between two significant quantities in the vacuum subtraction. 
Fortunately, in the calculation of $\alpha_E$, the existence of the factor
$t^2$ in the integral suppress the contributions from $H^{ii}(t)$ at small $t$. Nevertheless, the uncertainties for results from disconnected diagrams are still much larger than those from connected diagrams.
Finally, we obtain the disconnected contributions $\alpha_E^{ii,disc}$, which generally tend towards zero but carry uncertainties that are 2-3 times larger than those of the connected contributions.

\subsection{Quark-field contractions for $N\pi$ scattering and $N+\gamma^*\to N\pi$ transition}

\begin{figure*}[htb]
\centering
\includegraphics[width=0.24\textwidth,angle=0]{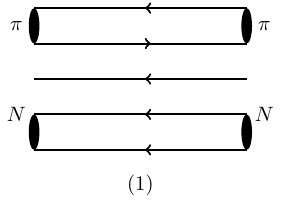}
\includegraphics[width=0.24\textwidth,angle=0]{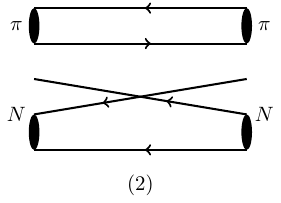}
\includegraphics[width=0.24\textwidth,angle=0]{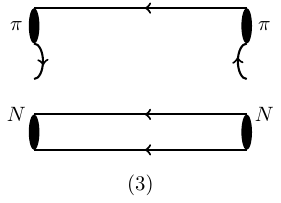}
\includegraphics[width=0.24\textwidth,angle=0]{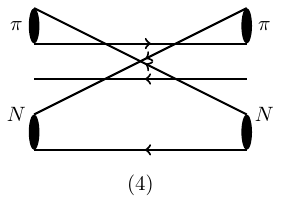}
\includegraphics[width=0.24\textwidth,angle=0]{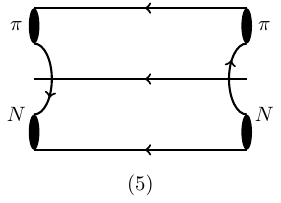}
\includegraphics[width=0.24\textwidth,angle=0]{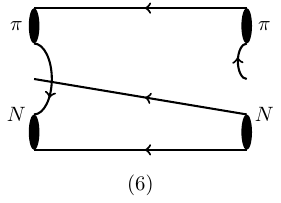}
\includegraphics[width=0.24\textwidth,angle=0]{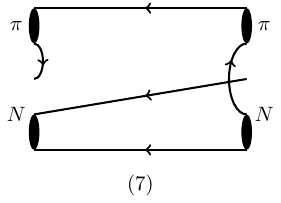}
\includegraphics[width=0.24\textwidth,angle=0]{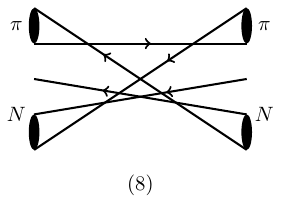}
\includegraphics[width=0.24\textwidth,angle=0]{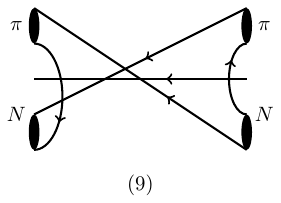}
\includegraphics[width=0.24\textwidth,angle=0]{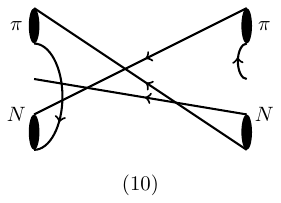}
\includegraphics[width=0.24\textwidth,angle=0]{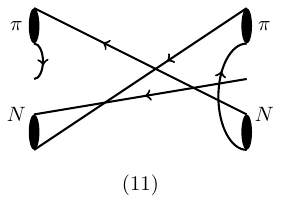}
\includegraphics[width=0.24\textwidth,angle=0]{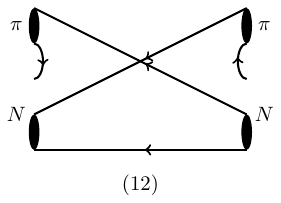}
\includegraphics[width=0.24\textwidth,angle=0]{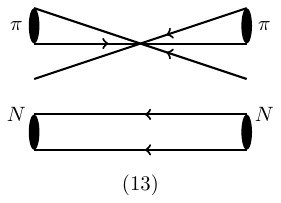}
\includegraphics[width=0.24\textwidth,angle=0]{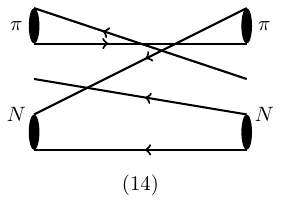}
\includegraphics[width=0.24\textwidth,angle=0]{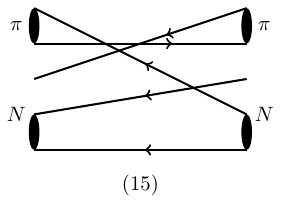}
\includegraphics[width=0.24\textwidth,angle=0]{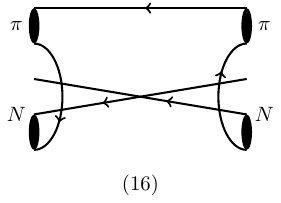}
\includegraphics[width=0.24\textwidth,angle=0]{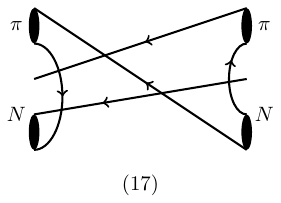}
\includegraphics[width=0.24\textwidth,angle=0]{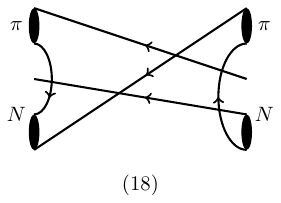}
\includegraphics[width=0.24\textwidth,angle=0]{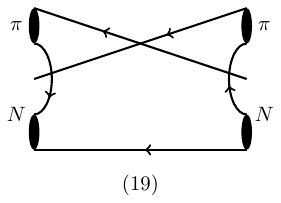}
\caption{Contractions for $N\pi$ scattering.
}
\label{fig:NpiNpi}
\end{figure*}

The $N\pi$ operators for different isospin channels are listed below
\ba
O_{N\pi}^{I=\frac{1}{2},I_3=\frac{1}{2}}&=&\frac{1}{\sqrt{3}}(\sqrt{2}O_nO_{\pi^+}+O_pO_{\pi^0}),
\nn\\
O_{N\pi}^{I=\frac{1}{2},I_3=-\frac{1}{2}}&=&\frac{1}{\sqrt{3}}(-O_nO_{\pi^0}+\sqrt{2}O_pO_{\pi^-}),
\nn\\
O_{N\pi}^{I=\frac{3}{2},I_3=\frac{1}{2}}&=&\frac{1}{\sqrt{3}}(O_nO_{\pi^+}-\sqrt{2}O_pO_{\pi^0}),
\nn\\
O_{N\pi}^{I=\frac{3}{2},I_3=-\frac{1}{2}}&=&\frac{1}{\sqrt{3}}(-\sqrt{2}O_nO_{\pi^0}-O_pO_{\pi^-}).
\ea
The correlation functions for the $N\pi$ scattering
\be
C_{N\pi}^I(t_f,t_i)=\langle O_{N\pi}^{I,I_3}(t_f) O_{N\pi}^{I,I_3,\dagger}(t_i)\rangle
\ee
encompass contributions from 19 distinct diagrams with 
\ba
C_{N\pi}^{I=\frac{1}{2}}&=&\frac{1}{2}\left\{2D_1-2D_2-3D_3
-2D_4-2D_5+3D_6+3D_7\right.
\nn\\
&&+3D_8+2D_9-3D_{10}-3D_{11}+3D_{12}+D_{13}
\nn\\
&&\left.-D_{14}-D_{15}-D_{16}+D_{17}+D_{18}-D_{19}\right\},
\nn\\
C_{N\pi}^{I=\frac{3}{2}}&=&D_1-D_2
-D_4-D_5+D_9-D_{13}+D_{14}+D_{15}
\nn\\
&&+D_{16}-D_{17}-D_{18}+D_{19}.
\ea
Here $D_i$ ($i=1,2\cdots,19$) designates the $i_{\mathrm{th}}$ type of contraction diagram, as illustrated in Fig.~\ref{fig:NpiNpi}.

\begin{figure*}[htb]
\centering
\includegraphics[width=0.24\textwidth,angle=0]{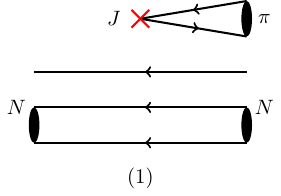}
\includegraphics[width=0.24\textwidth,angle=0]{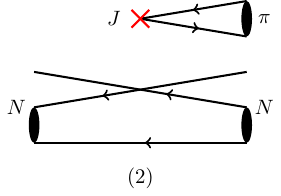}
\includegraphics[width=0.24\textwidth,angle=0]{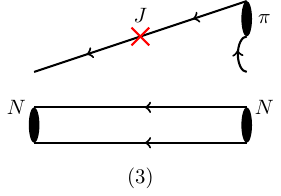}
\includegraphics[width=0.24\textwidth,angle=0]{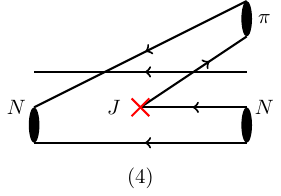}
\includegraphics[width=0.24\textwidth,angle=0]{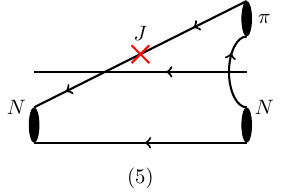}
\includegraphics[width=0.24\textwidth,angle=0]{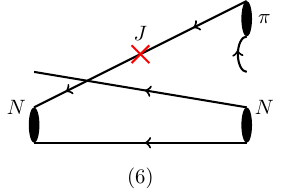}
\includegraphics[width=0.24\textwidth,angle=0]{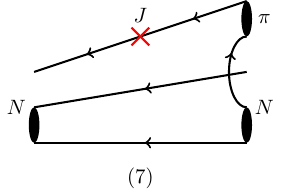}
\includegraphics[width=0.24\textwidth,angle=0]{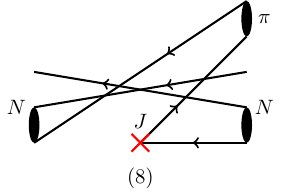}
\includegraphics[width=0.24\textwidth,angle=0]{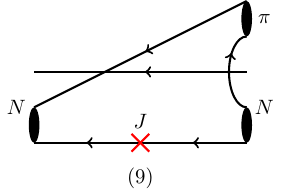}
\includegraphics[width=0.24\textwidth,angle=0]{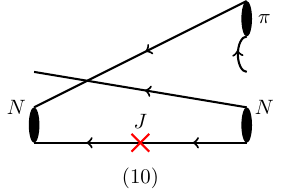}
\includegraphics[width=0.24\textwidth,angle=0]{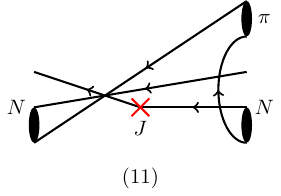}
\includegraphics[width=0.24\textwidth,angle=0]{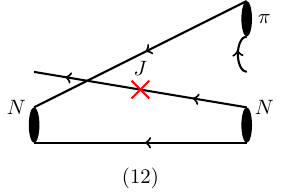}
\includegraphics[width=0.24\textwidth,angle=0]{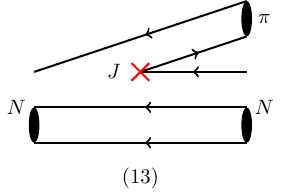}
\includegraphics[width=0.24\textwidth,angle=0]{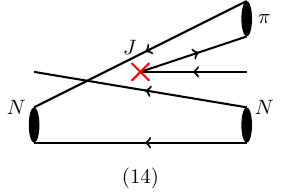}
\includegraphics[width=0.24\textwidth,angle=0]{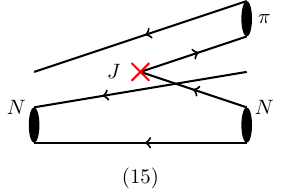}
\includegraphics[width=0.24\textwidth,angle=0]{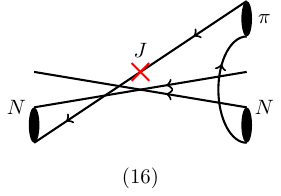}
\includegraphics[width=0.24\textwidth,angle=0]{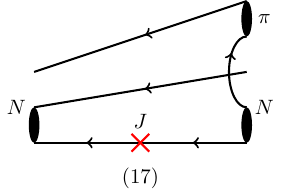}
\includegraphics[width=0.24\textwidth,angle=0]{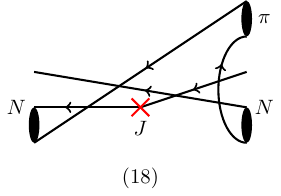}
\includegraphics[width=0.24\textwidth,angle=0]{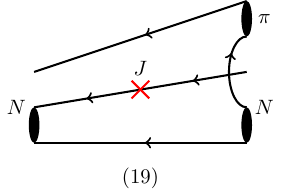}
\includegraphics[width=0.24\textwidth,angle=0]{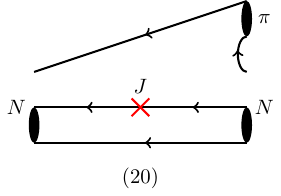}
\caption{Contractions for $N\gamma^*\to N\pi$ transition.
}
\label{fig:NJNpi}
\end{figure*}

The correlation functions relevant to the $N\gamma^*\to N\pi$ transition are given as
\be
C_{NJN\pi}^{I,I'}(t_f,t,t_i)=\langle O_{N\pi}^{I,I_3}(t_f) J_i^{I'}(t) O_{N}^\dagger(t_i)\rangle
\ee
where $I'$ signifies the isospin carried by the vector current, while $I$ and $I_3$ denote the isospin attributes of the $N\pi$ state.
These correlation functions incorporate contributions from $20$ diagrams with
\ba
C_{pJN\pi}^{\frac{1}{2},0}&=&-i\frac{\sqrt{2}}{4}
\left\{-E_3+E_6+E_7-E_8+E_{10}-E_{11}\right.
\nn\\
&&+E_{12}-E_{13}+E_{14}+E_{15}-E_{16}+E_{17}
\nn\\
&&\left.-E_{18}+E_{19}-2E_{20}\right\},
\nn\\
C_{pJN\pi}^{\frac{1}{2},1}&=&-i\frac{\sqrt{2}}{4}\left\{
2E_1-2E_2-3E_3-2E_4-2E_5+3E_6\right.
\nn\\
&&+3E_7+3E_8+2E_9-3E_{10}-3E_{11}+3E_{12}
\nn\\
&&\left.+E_{13}-E_{14}-E_{15}-E_{16}+E_{17}+E_{18}-E_{19}\right\},
\nn\\
C_{pJN\pi}^{\frac{3}{2},1}&=&i\left\{
E_1-E_2-E_4-E_5+E_9-E_{13}+E_{14}\right.
\nn\\
&&\left.+E_{15}+E_{16}-E_{17}-E_{18}+E_{19}\right\}.
\ea
Here $E_i$ ($i,=1,2,\cdots,20$) denotes the $i_{\mathrm{th}}$ type of contraction diagram, as depicted in Fig.~\ref{fig:NJNpi}.
When substituting the neutron for the proton in the initial state, the pertinent correlation functions are directly related to those in the proton case, as shown below
\be
C_{nJN\pi}^{\frac{1}{2},0}=C_{pJN\pi}^{\frac{1}{2},0},\quad C_{nJN\pi}^{\frac{1}{2},1}=-C_{pJN\pi}^{\frac{1}{2},1},\quad C_{nJN\pi}^{\frac{3}{2},1}=C_{pJN\pi}^{\frac{3}{2},1}.
\ee

\subsection{Matrix elements for low-lying $N\pi$ states}

\begin{figure*}[htb]
\centering
\includegraphics[width=0.49\textwidth,angle=0]{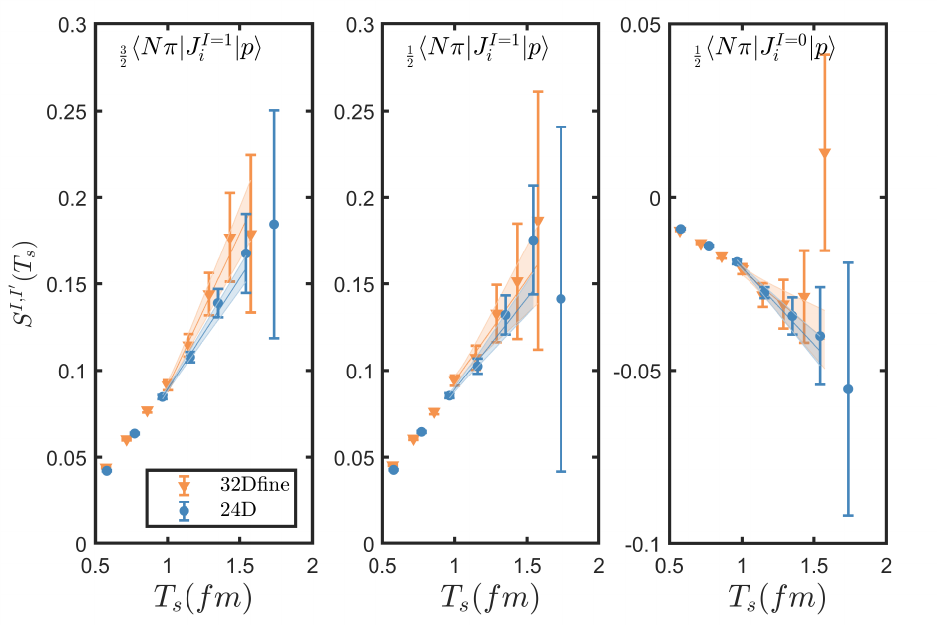}
\includegraphics[width=0.49\textwidth,angle=0]{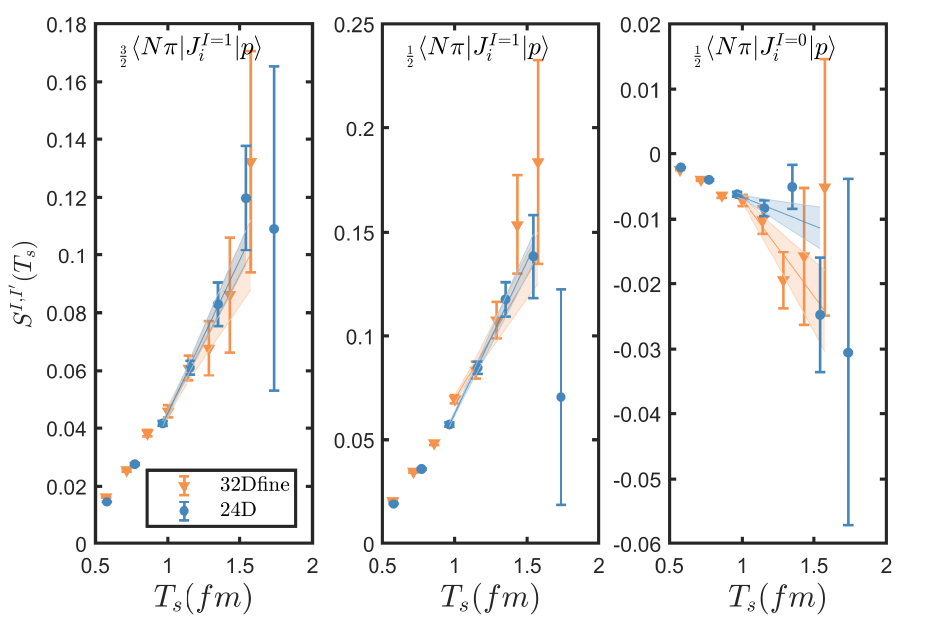}
\includegraphics[width=0.49\textwidth,angle=0]{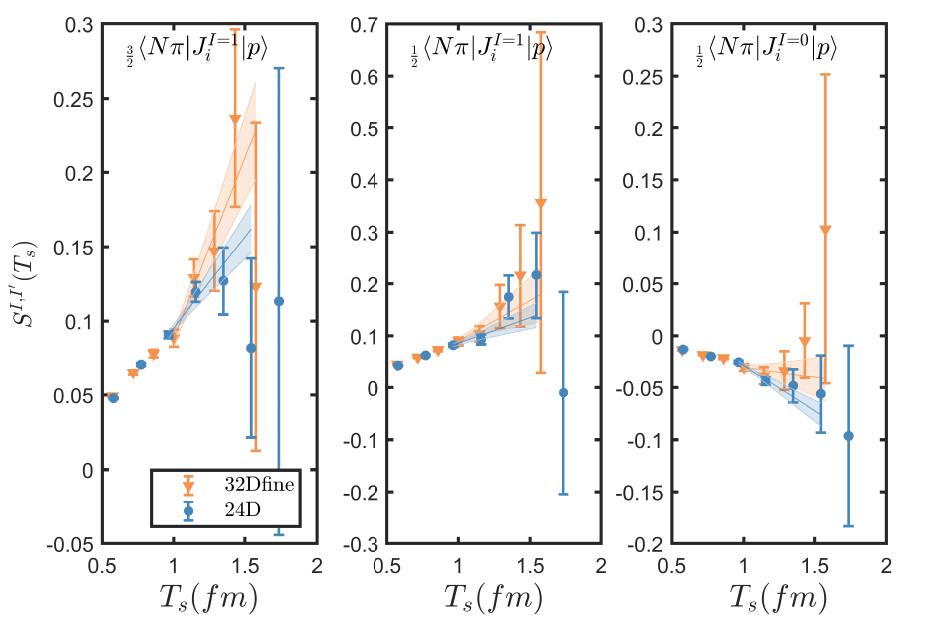}
\includegraphics[width=0.49\textwidth,angle=0]{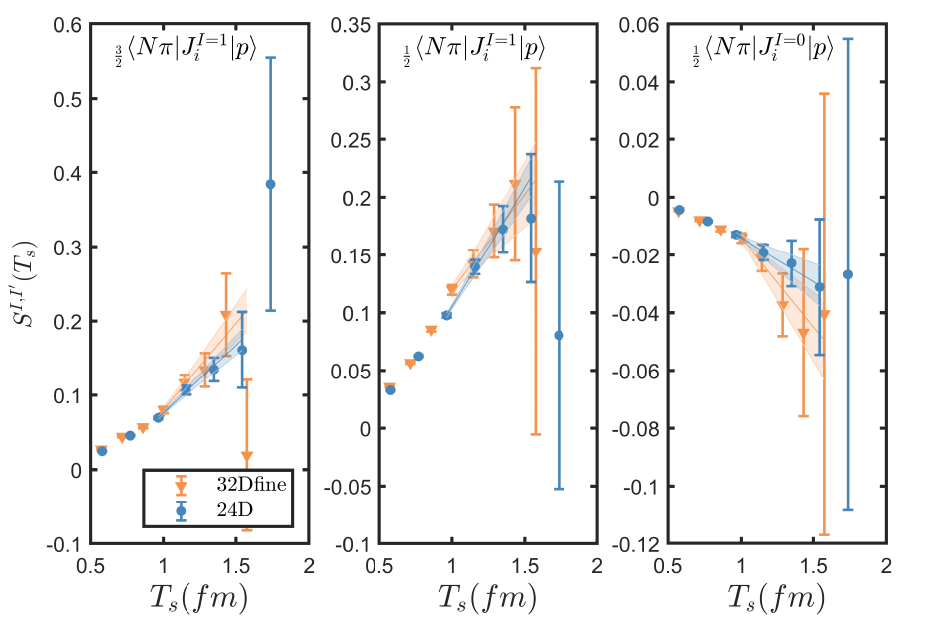}
\includegraphics[width=0.49\textwidth,angle=0]{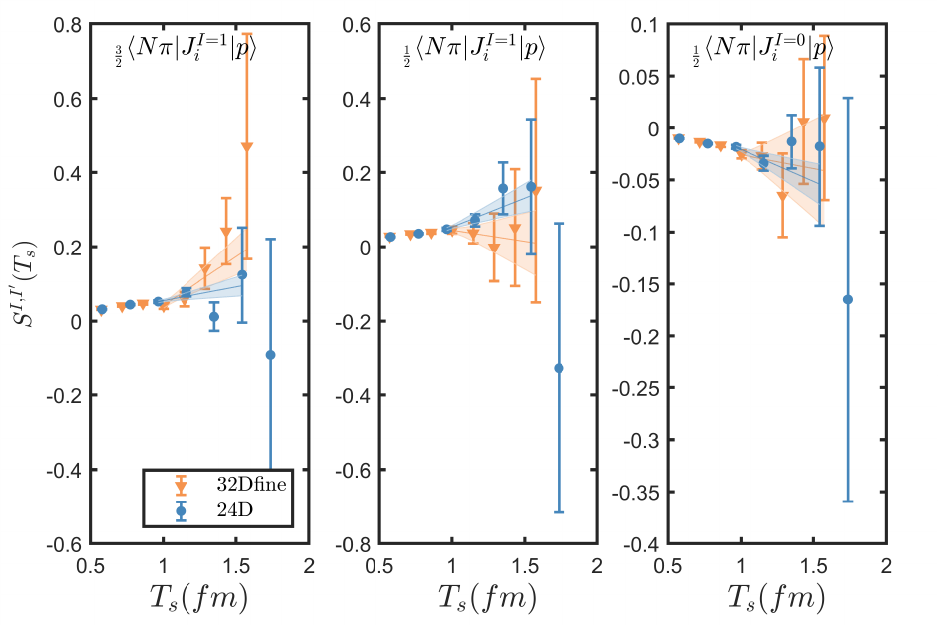}
\includegraphics[width=0.49\textwidth,angle=0]{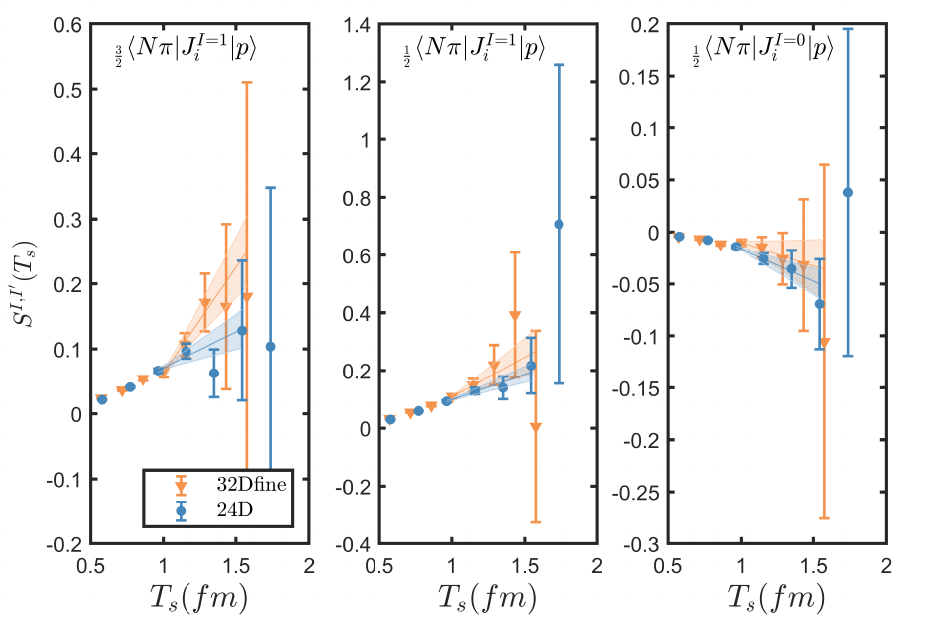}
\caption{The summed insertion $S^{I,I'}(T_s)$ as a function of $T_s$. The lattice results for the  $G_1^-$ and $H^-$ representation are presented on the left-hand side and right-hand side, respectively. From top to bottom, the results correspond to momentum modes $\vec{p}=\frac{2\pi}{L}\vec{m}$ with
$|\vec{m}|=1,2,3$.
}
\label{fig:matrix_elements_various_states}
\end{figure*}

The summed insertion $S^{I,I'}(T_s)$ for the ground state in the $G_1^-$ representation has been shown by Fig.~\ref{fig:summation_method}. In Fig.~\ref{fig:matrix_elements_various_states} we present results for the states corresponding to the momentum modes $\vec{p}=\frac{2\pi}{L}\vec{m}$ with
$|\vec{m}|=1,2,3$ for both $G_1^-$ and $H^-$ representations. By fitting $S^{I,I'}(T_s)$ to a linear form of $T_s$, we extract the matrix elements $\frac{1}{\sqrt{2M}}{_I}\langle N\pi|J_i^{I'}|N\rangle$.

\end{document}